\documentclass{aa}  

\usepackage{graphicx}
\usepackage{txfonts}
\usepackage{natbib}
\bibpunct{(}{)}{;}{a}{}{,}
\begin{document}

   \title{Revealing impacts of stellar mass and environment on galaxy quenching}

   \subtitle{}

   \author{Zhiying Mao
          \inst{1,2}
          \and
          Tadayuki Kodama\inst{1}
          \and
          Jose Manuel Pérez-Martínez \inst{1}
          \and
          Tomoko L. Suzuki \inst{3}
          \and
          Naoaki Yamamoto \inst{1,2}
          \and
          Kouta Adachi \inst{1}
          }

   \institute{Astronomical Institute, Graduate School of Science, Tohoku University, 6--3 Aoba, Sendai 980-8578, Japan\\
              \email{mzyirene@astr.tohoku.ac.jp}
         \and
             Graduate Program on Physics for the Universe (GP-PU), Tohoku University, 6--3 Aoba, Sendai 980-8578, Japan
         \and
            Kalvi Institute for the Physics and Methematics of the Universe, University of Tokyo, 5-1-5 Kashiwanoha, Kashiwa, Chiba, 277-8583, Japan
             }

   \date{}

 
  \abstract
   {}
   {{Galaxy quenching is a critical step in galaxy evolution. In this work, we present a statistical study of galaxy quenching in 17 cluster candidates at 0.5<z<1.0 in the COSMOS field.}}
   {{We selected cluster members with a wide range of stellar mass and environment to study their mass and environment dependence. Member galaxies are classified into star-forming, quiescent and recently-quenched galaxies (RQG) using the rest-frame UVJ diagram. We further separated fast and slow quenching RQGs by model evolutionary tracks on the UVJ diagram. We defined the quenching efficiency as the ratio of RQGs over star-forming galaxies and the quenching stage as the ratio of RQGs over quiescent galaxies to quantify the quenching processes.}}
   {{We found quenching efficiency is enhanced by both higher stellar mass and denser environment. Massive or dense environment galaxies quench earlier. Slow quenching is more dominant for massive galaxies and at lower redshifts, but no clear dependence on the environment is found. Our results suggest that low-mass galaxies in dense environments are likely quenched through a short-timescale process such as ram pressure stripping, while massive galaxies in a sparse environment are mostly quenched by a longer-timescale process. Using the line strength of H$\delta$ and [OII], we confirmed that our UVJ method to select RQGs agrees with high S/N DEIMOS spectra. However, we caution that the visibility time (duration of a galaxy's stay in the RQG region on the UVJ diagram) may also depend on mass or environment. The method introduced in this work can be applied to RQG candidates for future statistical RQG spectroscopic surveys. The systematic spectroscopic RQG study will disentangle the degeneracy between visibility time and quenching properties.}}
   {}


   \maketitle
%

\section{Introduction}
{Deep surveys have transformed our view of the distant Universe, allowing us to constrain the process of galaxy formation and evolution over the past 13 billion years of cosmic history. The cosmic star formation rate density comes to a peak at $z\sim2-3$, and then it declines as the Universe ages until the present day \citep{Lilly_1996,Karim_2011,Burgarella_2013,Madau_2014,Khostovan_2015}. Such decline in star formation is called galaxy quenching, an essential step in galaxy evolution. Although there have been many efforts to understand quenching mechanisms, there is still a lack of conclusive results.

Quenching is a process where star-forming galaxies (SFGs) transform into quiescent galaxies (QGs). The transitional population between SFGs and QGs are RQGs, which are ideal for studying the quenching process \citep{Wu_2020}. The recently-quenched galaxies are still in or shortly after the quenching process \citep{Newberry_1990,Wild_2009,French_2018}. This nature makes them capable of delivering information about the physical processes during quenching. RQGs have quenched their star formation in the past few tens million years to a few billion years\citep{Couch_1987,Poggianti_2009}, but they still contain a large fraction of relatively young A-type stars. This particular stellar composition adds strong Balmer absorption lines to the passive galaxy spectrum \citep{Dressler_1983,Couch_1987,Wild_2009}. In previous studies, this population is mostly referred to as post-starburst galaxies (PSBs), which means galaxies that have experienced a starburst episode followed by a fast quenching. However, this starburst phase is not essential for quenching, and the quenching timescale varies for each galaxy \citep{Poggianti_2009}. Therefore, we use "recently-quenched galaxy" in this work as a more precise name for this population.

There are two key parameters to describe galaxy quenching, mass, and environment, as the red fraction of galaxies is known to correlate strongly with these two parameters  \citep{Peng_2010,Sobral_2011,Muzzin_2012,Darvish_2016}. Physical mechanisms of quenching are correspondingly categorized into mass-dependent and environment-dependent types.

Among the stellar mass driven quenching, the most studied mechanism is AGN feedback \citep{Best_2005,Hopkins_2005,Kaviraj_2007,Diamond_2012,Belli_2021}, which is also studied in simulations (e.g. \citealt{Dubois_2013,Beckmann_2017,Weinberger_2017,Piotrowska_2021,Shi_2022,Xu_2022}). The quasar host AGNs show evidence of having strong outflows in observation\citep{Dai_2008,Tombesi_2010}. AGN outflows cause expulsion of the gas from the gravitational bound of the system and hence reduce the subsequent star formation activity in their host galaxies \citep{Debuhr_2012,Combes_2017}. AGNs can also heat up the cool gas of the interstellar medium, hence preventing the galaxy from forming new stars \citep{Croton_2006,Man_2018,Zinger_2020}. Supernovae feedback is another primary mechanism of mass-dependent quenching that mainly acts on low-mass galaxies. High energy ejections and superwinds take away the cold gas, stopping further star formation activity \citep{Dekel_1986,Dekel_2003}. 

On the other hand, quenching mechanisms related to interactions between galaxies and their environment (e.g. other galaxies, intracluster medium (ICM), intergalactic medium (IGM)) are considered environmental quenching. A widely established mechanism of environmental quenching is strangulation, also known as 'starvation' \citep{Larson_1980}. The virial shocks from the halo heat up the cool gas, ceasing galaxies' gas supply. The star formation will then be quenched due to the lack of fuel \citep{Bekki_2002,Man_2018}. Galaxy mergers can be another mechanism of environmental quenching. Especially, major mergers can trigger starbursts which are followed by rapid consumption of all remaining cool gas \citep{Mihos_1996,Gabor_2010,Man_2018}. In addition, when galaxies encounter at high speed, their potential wells will interact with each other, causing gravitational shocks. Even though the effect of individual encounters may not be large, the repeated process in high-density environments would make a significant impact on galaxies. This process is named as galaxy harassment \citep{Moore_1996,Moore_1998}. Galaxy harassment leads to instability of galaxy morphology and change in the distribution of material, driving gas into central regions \citep{Moore_1999,Boselli_2006}. Galaxies will be quenched after a starburst consumes their gas. Ram pressure stripping is also an effective environmental quenching mechanism \citep{Ma_2008,Bekki_2009,Roediger_2014}. As galaxies move through ICM in cluster cores, the 'wind' from ICM will strip off the gas from them \citep{Gunn_1972,Balogh_2000,Steinhauser_2016}. 

Many mechanisms that probably lead to quenching have been proposed, and surely we need some parameters to discriminate these mechanisms. One of such critical parameters we are looking for is the quenching timescale, which varies from one mechanism to another. Most of these timescales are obtained from simulations \citep{Wright_2019, Wetzel_2013,Walters_2022}. Among mass-dependent quenching mechanisms, the supernova feedback is a strong physical process that quenches star formation fast in $\sim$0.1 Gyr \citep{Ceverino_2009}. On the other hand, the AGN feedback timescale is still under debate. Quasar mode AGN feedback is a strong process. Quasars' lifetimes are not long (<0.1 Gyr, \citealt{Hopkins_2006}), and the gas outflow driven by quasar quenches star formation in million-year-level timescales \citep{Smethurst_2021}. On the other hand, the radio mode of AGN feedback is a slow process \citep{Best_2005}; it takes up to and beyond 1 Gyr to establish a balance between cooling and heating to reach a low star formation rate phase\citep{Fabian_2012}. In \citet{Schawinski_2014}, the authors discuss that the timescale of AGN feedback leading to a quenching can be related to the host galaxy type. Furthermore, \citet{Hirschmann_2017} simulated star formation history (SFH) of galaxies with AGN and found that the quenching timescale can have a wide range, from a few hundred Myr to a few Gyr. For environmental quenching mechanisms, strangulation is a long-term process, which lasts for a few billion years \citep{Peng_2015}. Merger-driven quenching does not happen directly after galaxy merger events, a median delay time of 1.5 Gyr is expected, and the timescale varies over a wide range \citep{Rodriguez_2019}. Ram pressure stripping is a rapid type of quenching, with a timescale of $\sim$0.2 Gyr \citep{Steinhauser_2016}. Thus, we can use the quenching timescale to distinguish some physical mechanisms involved.

To study the quenching process, the main drawback of current RQG works is the lack of observations. Due to their intrinsic short-lived nature, RQGs are rare objects \citep{Tran_2004,LeBorgne_2006,Kaviraj_2007}, and we still lack a large spectroscopic sample to carry out systematic studies. Although RQG spectra can provide direct information and strict constraints on their physical properties and SFH, only a limited number of RQG spectra exist at high redshifts (e.g. \citealt{Pracy_2010,Maltby_2016,D'Eugenio_2020}). Increasing the spectroscopic sample is critical, and for that purpose, we first need to make a large photometric sample of RQG candidates to feed spectroscopic follow-up observations. Moreover, such a statistical photometrically selected sample of RQG candidates is essential for us to explore the quenching processes prior to spectroscopic studies. As the quenching process is strongly correlated with mass and environment, a systematic study of RQGs must cover a wide range of masses and environments. Therefore we require a sample of RQGs extending from cluster cores to general fields in order to understand environmental quenching. There are some previous works which study the environment-dependent quenching using the post-starburst galaxy (PSB) population, which belongs to the RQG population (e.g. \citealt{Socolovsky_2018, Paccagnella_2017, Paccagnella_2019}). They study the environment-dependent quenching by estimating PSB fraction or number in different environments (e.g. cluster, group, or field; distance to cluster centre). These studies prove that the PSBs tend to reside in denser environments. Their results also suggest that there should be various quenching mechanisms in different environments \citep{Paccagnella_2017}. In clusters, fast quenching processes, such as ram pressure stripping, are enhanced. In sparse environments, galaxy mergers and interactions are producing PSBs \citep{Socolovsky_2018,Paccagnella_2019}. 
There are also works on the mass dependence of galaxy quenching. \citet{McNab_2021} studied the mass dependence of quenching for transitional populations (similar to RQGs) using spectroscopic data from the GOGREEN survey \citep{Balogh_2017}. Their results support the scenario of pre-processing in the group or proto-cluster environment for massive galaxies and the fast quenching during infall for lower mass galaxies. The GOGREEN survey contains spectra of many cluster galaxies with secure redshift at $0.8<z<1.5$, which will improve the study of transitional populations in the future.

In this work, we present a systematic study of recently-quenched galaxies (RQGs) based on multi-band photometric data in 17 cluster candidates in the COSMOS field at $0.5<z<1.0$. We aim to develop an efficient method to select RQG from photometric surveys (e.g. HSC-SSP, \citealt{Aihara_2019}) to prepare for the coming spectroscopic RQG survey using Prime Focus Spectrograph (PFS, \citealt{Tamura_2018}) on Subaru. We select our RQG sample with the rest-frame UVJ colour-colour diagram and confirm some of the proposed quenching mechanisms using this sample. With the current photometrically-selected RQGs, we conduct basic analyses on mass and environment dependence of galaxy quenching and get some preliminary results. This paper is structured as follows. In Sect. \ref{sec:data}, we present the photometric data we use and the cluster galaxy sample we selected. In Sect. \ref{sec:method}, we present the method we use to classify RQGs and correct for possible projection effects. In Sect. \ref{sec:result}, we display our results of quenching efficiency, quenching stage, and quenching timescale. In Sect. \ref{sec:discussion}, we discuss the possible contamination and incompleteness of our RQG selection, compare our results with previous works, and then attempt to interpret the results we obtain from the current sample. An $H_0 = 69.6 {\rm km s^{-1} Mpc}$, $\Omega_M = 0.286$ and $\Omega_{\Lambda}=0.714$ cosmology is used. All cosmology calculation is conducted by the cosmology calculator from \citet{Wright_2006}. All magnitudes presented in this paper are defined in the AB magnitude system.}

\section{Dataset and sample}\label{sec:data}
In Sect. \ref{sec:photometric}, we describe the photometric data we use in this study. In Sect. \ref{sec:sample}, we describe how we select our sample galaxies to study the quenching processes.

\subsection{The COSMOS2015 catalogue}\label{sec:photometric}
The photometric data we use in this work is part of the COSMOS2015 catalogue \citep{Laigle_2016}. This catalogue mainly combines 26 bands from optical to infrared wavelength of galaxies in the COSMOS field. It offers deep ($m_{AB}\sim$25-26 mag) data covering a relatively large area of 2 ${\rm deg}^2$. As we require near-infrared (NIR) photometry to estimate rest-frame J band magnitude, the data we use is limited to the central $\sim1.4 {\rm deg}^2$ of the COSMOS field. In this work, we use $B, V, r, i^+, z^{++}$ taken with Suprime-Cam/Subaru, $Y, J, H, K_s$ bands taken with VIRCAM/VISTA, and $3.6\mu m, 4.5\mu m$ taken with IRAC/{\it Spitzer}. Broad band  $B, V, r, i^+$ data are taken from the previous data release \citep{Capak_2007,Ilbert_2009}. Broad $z^{++}$ band data are taken with new upgraded CCDs. For each optical band, image PSFs are homogenised so that tile-to-tile variations can reach minimisation \citep{Capak_2007}. NIR $Y, J, H, K_s$ band data are part of the UltraVISTA survey, and the data combined with the COSMOS2015 catalogue was released in UltraVISTA-DR2, which is much deeper in ultra-deep stripes compared to DR1 \citep{McCracken_2012}. The IRAC $3.6\mu m, 4.5\mu m$ data are a combination of several programs observing the COSMOS field using {\it Spitzer} \citep{Capak_2016}.
We correct for foreground galactic extinction according to the catalogue included reddening value EBV (=E(B$-$V)) for each object and foreground extinction factor $F_f$ \citep{Allen_1976} according to \citet{Laigle_2016}:
\begin{equation}
{\rm MAG}_{\rm i,f,corrected}={\rm MAG}_{i,f}-{\rm EBV}_i\times F_f
\end{equation}

\smallskip
In this work, we take photometric redshift and stellar mass estimation results from \citet{Laigle_2016}. In that work, spectral energy distribution (SED) fitting and photometric redshift estimation are performed with the LEPHARE code \citep{Arnouts_2002, Ilbert_2006}.
They use a set of 31 templates, including spiral and elliptical galaxies from \citet{Polletta_2007} together with 12 templates of young blue star-forming galaxies generated by \citet{BC03} models. Dust attenuation is added as a free parameter. 3" aperture magnitudes are adopted since they provide slightly better photometric redshift. The code performs $\chi^2$ analysis between template predicted and observed fluxes. Stellar masses are then derived by the method of \citet{Ilbert_2015}, using a spectral library generated by \citep{BC03} model. The \citet{Chabrier_2003} initial mass function is assumed. 

A 97\% confidence-level spectroscopic redshift sample is obtained from \citet{2007ApJS..172...70L} to examine the reliability of photometric redshift. The precision of photometric redshift is estimated using normalized median absolute deviation (NMAD, \citealt{Hoaglin_1983}), which is defined as 1.48 $\times$ median$(|z_p-z_s|/(1+z_s))$, where $z_p$ refers to photometric redshift and $z_s$ refers to spectroscopic redshift. The dispersion of NMAD of the spectroscopically confirmed sample is $\sigma$=0.021. Since the width of redshift slices we use in this study is 0.1, this $\sigma$ level is precise enough for our further study. This work focuses on the redshift range $z\sim0.5-1.0$, where we have more confirmation of photometric redshift and higher mass completeness (see \ref{sec:sample}). Moreover, in this range, the UVJ selection criteria are not affected by redshift; hence we can apply the same UVJ selection region for the whole sample (more information in Appendix \ref{app:model track}).

\subsection{Sample selection}\label{sec:sample}
In this section, we describe how we select galaxy cluster candidates and their member galaxies.

We followed the steps in \citet{Laigle_2016} to obtain mass completeness for this sample. 
For each galaxy, the mass required for it to be observed ($M_{lim}$) is determined based on $K_s$ magnitude and stellar mass M according to the following equation,
\begin{equation}
{\rm log}\,M_{lim} = {\rm log}\,M - 0.4(K_{s,lim}-K_s)    
\end{equation}
For each galaxy, the 5$\sigma$ limiting magnitude, $K_{\rm s,lim}$, is used to estimate $M_{\rm lim}$, the stellar mass it requires to be observed. Then in each redshift bin, the mass completeness is set at $M_{\rm lim}$ down to which 90\% of galaxies are contained. The qualitative conclusions we draw remain unchanged when we apply a slightly higher mass cut, such as 95\%. In the overall sample, we remove galaxies without a stellar mass estimation and adopt the most conservative mass limit in the highest redshift bin at $M_{\rm lim} = 10^{9.8}M_{\odot}$. All further analyses are carried out for galaxies with stellar masses higher than the mass limit based on the SED fitting. 

We then select galaxies with type flag=0 (galaxy type). Since we use five optical bands from Subaru and four UVista near-infrared bands to calculate rest-frame UVJ colour, we require our target galaxy to cover at least two optical and two NIR bands to ensure the accuracy of rest-frame colour estimation. After these selections, we have 31983 galaxies. We further select our analysis sample and field sample from these galaxies.

In order to study the environmental dependence of galaxy properties in an unbiased way, we select our sample galaxies from the cluster cores to the outskirts. Thus, our first step is to search for galaxy cluster candidates at $0.5<z<1$.

\subsubsection{Overdensity identification}
In this study, we select overdensity regions based on the aperture density method, computing the aperture density (i.e. the number density of galaxies in a certain aperture) across our field using 1 Mpc radius apertures. Since there is no complete cluster catalogue at this redshift range, we choose to identify our own cluster candidates to keep consistency throughout this work in the selection and analysis of our cluster and field samples.

\begin{figure}
	\includegraphics[width=\columnwidth]{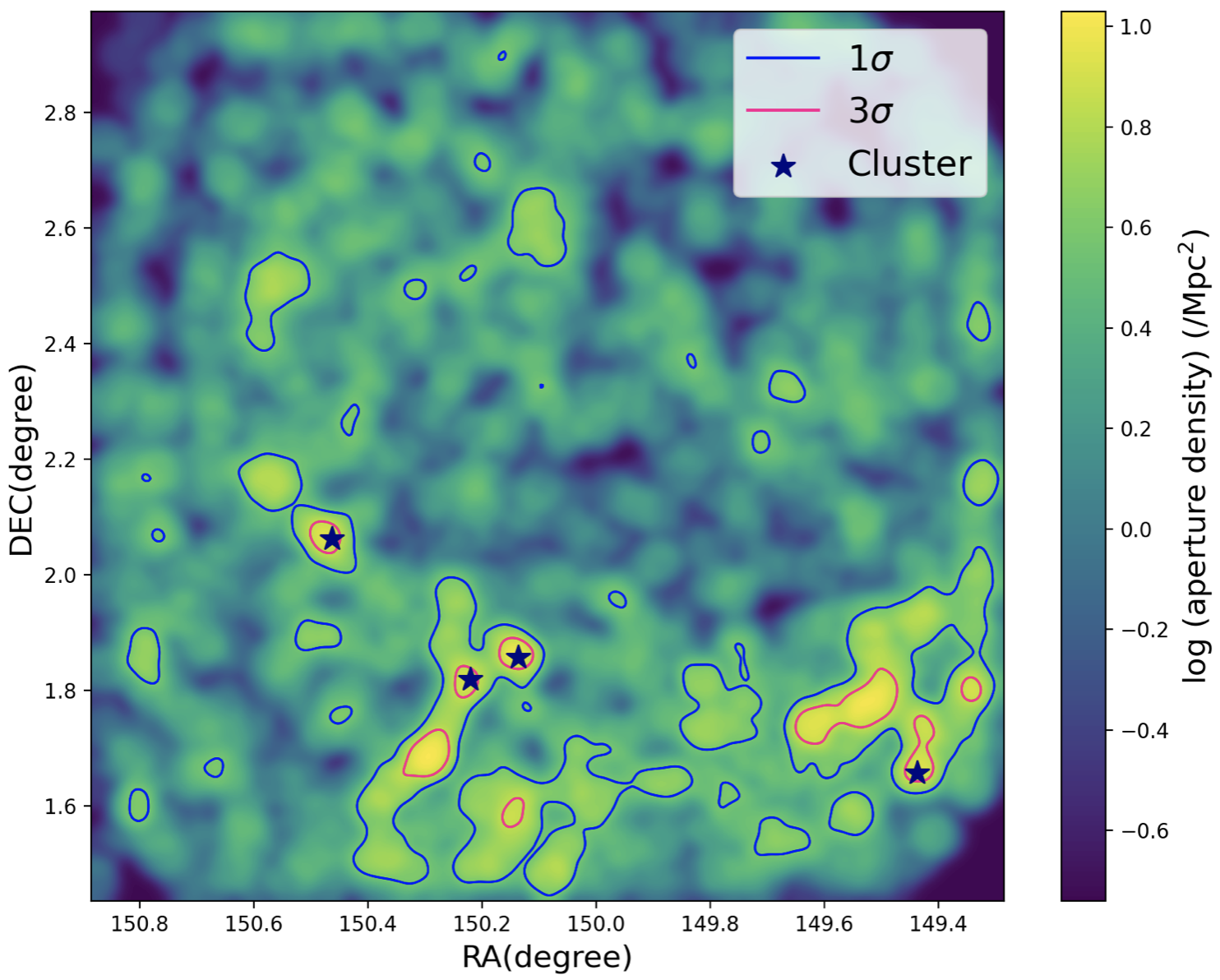}
    \caption{An aperture density map of a redshift slice at z$\sim$0.5--0.6. Colours show aperture density in the logarithm scale. Red contours mark the edge of 3$\sigma$ peaks, and the blue contours mark the edge of 1$\sigma$ regions. The area outside 1$\sigma$ contours is identified as field. This kind of density map is plotted for all redshift slices to select cluster candidates and construct the field sample, which is used to make a statistical subtraction of the field average density. Stars mark confirmed cluster candidates we analyse in this work.}
    \label{fig:density map}
\end{figure}

We first slice the overall sample into 5 redshift bins with a width of $\Delta z= 0.1$ each. Then we calculate the aperture density distribution for each slice and make the aperture density maps (e.g. Fig. \ref{fig:density map}) by placing 1 Mpc radius apertures on 1 Mpc $\times$ 1 Mpc grids that cover the whole COSMOS field. We adopt this aperture size to capture overdensity on the cluster scale, since 1Mpc is similar to the radius of a $10^{14}M_{\odot}$ level cluster \citep{Chiang_2013}. Then we define $\sigma$ as the scatter of densities in all apertures. Regions with $>3\sigma$ overdensity (i.e. density is at least $3\sigma$ higher than average density) are selected as possible cluster candidates. In some cases, the redshift distribution of cluster members is too close to the edge of the redshift slice under scrutiny. Thus, we shift the redshift bin by z$\sim$0.05 to include the majority of the members (see more in Sect. \ref{sec:z confirmation}). We selected 37 overdensity regions in all redshift bins. 8 of them are the same overdensity regions at different redshifts. As a result, we found 29 different overdensity regions in $0.5<z<1.0$.

\begin{table}
\centering
\caption{Information of cluster candidates studied in this work {(accepted clusters in Sect. \ref{sec:z confirmation})}. Column one shows the ID of each candidate; column two shows the selection redshift slice they are in; column three and four present the coordinate of the cluster candidate centre (RA and DEC); column five presents the mean photometric redshift value of all galaxy members, and the last column shows the scatter of cluster candidate members' redshift distribution.}
\begin{tabular}{cccccc}
\hline
ID & Redshift slice & RA & DEC &Mean & $\sigma_z$ \\
& &[degree] &[degree] & redshift & \\
\hline
01 & 0.5-0.6 & 150.463 & 2.062 & 0.548 & 0.0199\\
02 & 0.5-0.6 & 150.210 & 1.819 & 0.538 & 0.0234\\
03 & 0.5-0.6 & 150.136 & 1.847 & 0.542 & 0.0245\\
04 & 0.5-0.6 & 149.437 & 1.655 & 0.540 & 0.0228\\
05 & 0.55-0.65 & 150.579 & 2.473 & 0.600 & 0.0244\\
06 & 0.6-0.7 & 150.447 & 1.882 & 0.662 & 0.0238\\
07 & 0.65-0.75 & 149.922 & 2.525 & 0.707 & 0.0216\\
08 & 0.65-0.75 & 150.167 & 2.523 & 0.685 & 0.0229\\
09 & 0.8-0.9 & 150.445 & 2.106 & 0.851 & 0.0234\\
10 & 0.8-0.9 & 150.504 & 2.225 & 0.838 & 0.0213\\
11 & 0.8-0.9 & 150.417 & 1.977 & 0.845 & 0.0229\\
12 & 0.8-0.9 & 149.623 & 2.400 & 0.840 & 0.0216\\
13 & 0.85-0.95 & 149.962 & 2.650 & 0.894 & 0.0238\\
14 & 0.85-0.95 & 150.083 & 2.535 & 0.891 & 0.0233\\
15 & 0.85-0.95 & 150.163 & 2.597 & 0.899 & 0.0245\\
16 & 0.9-1.0 & 150.093 & 2.199 & 0.944 & 0.0241\\
17 & 0.9-1.0 & 149.985 & 2.317 & 0.938 & 0.0187\\
\hline\\
\end{tabular}
\label{tab:cluster sample}
\end{table}

\subsubsection{Member galaxy selection}
We have adopted 1 Mpc radius apertures to select overdensities, but they cannot trace fine structures and local environments within cluster candidates. Thus, we use the local density $\Sigma_5$ to probe the local environment. $\Sigma_5$ is defined as the average density within a circle centred at a given position with a radius of the distance between the central point and its $5_{\rm th}$ nearest neighbour. We note that we should not use average $\Sigma_5$ as the overall field average density due to the uncertainty of photometric redshift. In redshift confirmed overdensities (see more in Sect. \ref{sec:z confirmation}), the nearest neighbours have lower probabilities of being a foreground or background galaxy. However, in low-density field areas, the impact of projection effects is strong and cannot be corrected. To determine the centre of the cluster candidates, the $\Sigma_{5}$ values are calculated on 0.01 Mpc steps covering the 1 Mpc $\times$ 1 Mpc square region centring the peak of aperture density in each cluster candidate. The $\Sigma_{5}$ peaks within the mentioned square region are determined as the centres. The information of these cluster candidate centres is displayed in Table \ref{tab:cluster sample}.

We then put a square aperture at the cluster candidate's centres and select all galaxies in this redshift slice to be members of this cluster candidate. The actual cluster size depends on mass, redshift and cluster model. Here we apply fixed 2.8 Mpc $\times$ 2.8 Mpc apertures for simplicity because most of our clusters do not have a measured virial radius. We choose a relatively large aperture size to capture more members of clusters, especially massive clusters. The aperture size we choose is still consistent with the typical cluster radius $\sim$ 1.1 - 1.6 Mpc at $z\sim 0.8$ \citep{Chiang_2013}. The field galaxies will not be huge contamination because we check the redshift distribution of cluster members (see Sect. \ref{sec:z confirmation}) and subtract field galaxies to correct for the projection effect (see Sect. \ref{sec:subtraction}).
Since we discuss the environmental effect with local density $\Sigma_{5}$, the square aperture selection (instead of careful cluster member identification) is good enough for our analysis. Small changes in the dimensions (2.5 - 3 Mpc) and the shape of the aperture do not qualitatively change our results. We checked the overlap between cluster candidate regions. There are two cases: cluster candidates 02 and 03 have an overlap of 2.3 ${\rm Mpc}^2$ and cluster candidates 14 and 15 have an overlap of 0.56 ${\rm Mpc}^2$. We do not apply any special treatment for the overlapping regions; galaxies in that region are calculated twice as a member of each cluster candidate. Considering that the overlapping region is small, it does not affect our results.

\subsubsection{Redshift distribution confirmation}\label{sec:z confirmation}
For overdensity estimation, we included all galaxies within $\Delta z= 0.1$, which is larger than the real cluster redshift range. However, the photometric redshift has some uncertainty, and we do not have kinematic information on galaxies. Therefore, it could be possible that some of these overdensities are created by projection effects. Thus, we must confirm the distribution of the potential cluster candidate members in the redshift space. 
\begin{figure}
	\includegraphics[width=\columnwidth]{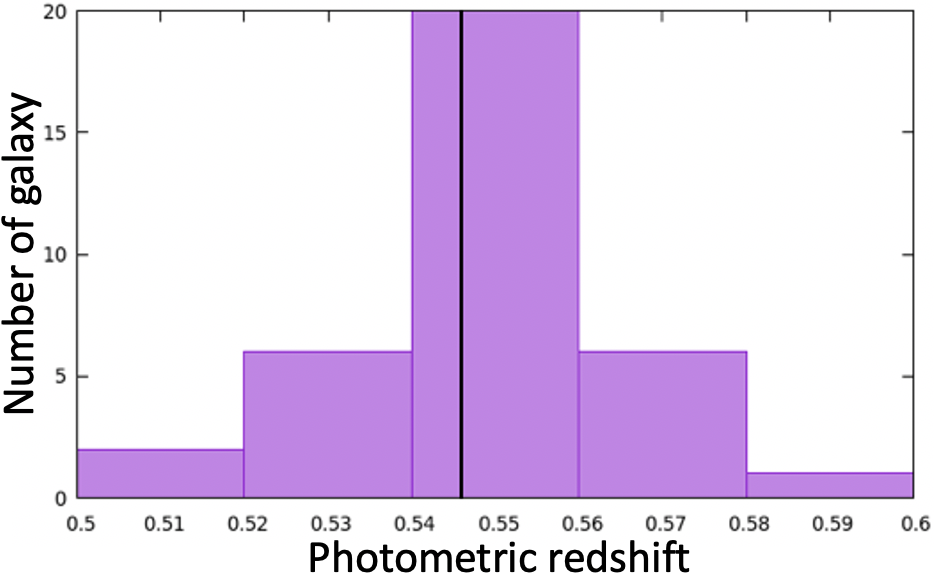}
	\includegraphics[width=\columnwidth]{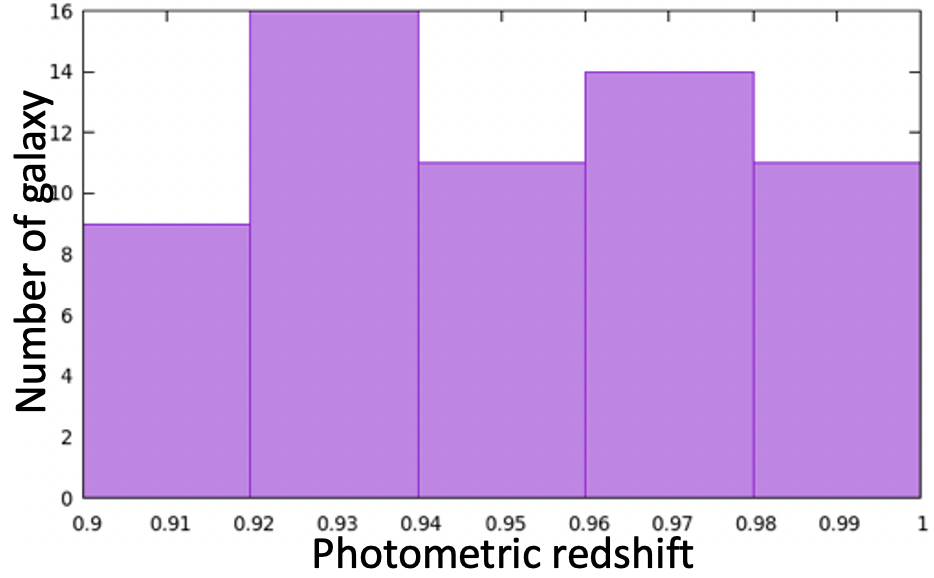}
	\includegraphics[width=\columnwidth]{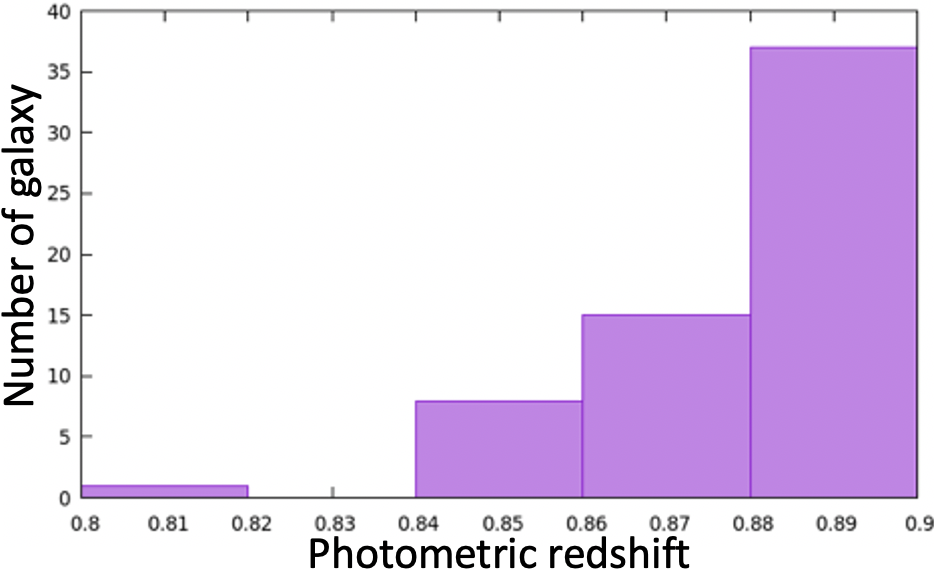}
    \caption{Redshift distribution of three typical cases. Histograms show the distribution of photometric redshifts of galaxy members. The top panel displays a cluster candidate located within one redshift slice; the vertical line in this panel gives the mean photometric redshift value. The middle panel shows a flat distribution, indicating that this overdensity is not a physically connected structure. The bottom panel shows the distribution of a real structure, but its peak is close to the edge of our selection redshift slice, missing a large fraction of its galaxy members.}
    \label{fig:z distribution}
\end{figure}

A real cluster should have Gaussian-like photometric redshift distribution (top panel of Fig. \ref{fig:z distribution}), while fake clusters, due to the projection effect, have a flat or a multi-peak distribution (middle panel of Fig. \ref{fig:z distribution}). To remove contamination from "fake" clusters, we calculate the scatter $\sigma_z$ of the photometric redshift distribution of cluster candidate members. $\sigma_z<0.025$ is required for an overdensity to be selected as a cluster candidate. In some other cases, redshift distribution shows one side of Gaussian with a peak towards the edge of the redshift bin (bottom panel of Fig. \ref{fig:z distribution}). This type of redshift distribution may belong to a real cluster at a different redshift. When $|\overline{z}-z_{\rm slice\;edge}|<0.03$, we shift the redshift bin by z$\sim$0.05, then check their overdensity and galaxy members' redshift distribution to determine if they belong to the final cluster candidate sample. All 17 accepted cluster candidates' redshift information is listed in Table \ref{tab:cluster sample}. Finally, we have a galaxy sample of 1396 galaxies.

\subsubsection{Comparison with other cluster catalogues}\label{sec:cluster catalogue}
We compare the final cluster candidates sample with several other cluster/group catalogues following different selection methods (e.g. overdensity, weak lensing, cosmic web structure). First, we compare it with Chandra COSMOS X-ray catalogue \citep{Gozaliasl_2019}. This work used extended X-ray sources detected on a 32-128 arcsec scale. The authors searched for the cluster/group members within the extended X-ray source area. And the redshift is mainly decided from spectroscopic redshift in \citet{Hasinger_2018}. We find 3 matches out of 14 clusters/groups in \citet{Gozaliasl_2019}. Cluster candidates 02 and 06 in our sample are matched with clusters 20035 and 10052 in \citet{Gozaliasl_2019} respectively within 0.03 Mpc. Cluster candidate 11 in our sample can match with cluster 20069 in \citet{Gozaliasl_2019} within 0.13 Mpc. The information of these three clusters is listed in Table \ref{tab:X-ray cluster}. 
Apart from the three cluster candidates mentioned above, another cluster in \citet{Gozaliasl_2019} can be found by our method, but it does not have NIR so we do not put it in our cluster candidate catalogue; two others are selected by overdensity mapping but ruled out due to flat photometric redshift distribution. Considering the size of the extended X-ray source is much smaller than the structure we are looking for, we also test using the smaller 0.5 Mpc radius aperture to search for overdensities again and reproduce almost all clusters/groups in \citet{Gozaliasl_2019} except the least massive one. However, these smaller size structures are more likely to be groups with small masses ($<$10$^{14}$M$_\odot$) rather than clusters, so we do not use them for our analysis.

\begin{table*}
    \centering
    \caption{Properties of X-ray detected clusters taken from \citet{Gozaliasl_2019}. In the first column, brackets are our own cluster IDs listed in Table \ref{tab:cluster sample}. The last column shows the distance between the X-ray detection and our definition of the cluster centre.}
    \begin{tabular}{cccccc}
         \hline
         ID & RA & DEC & ${\rm z}_{\rm spec}$ & ${\rm M}_{200}$ & Deviation \\
         &[degree]&[degree]&&[$\times {\rm 10}^{12}{\rm M}_{\odot}$]& [${\rm Mpc}$]\\
         \hline
         20035 (02)&150.210&1.820&0.529&47.09$\pm$4.38&0.02\\
         10052 (06)&150.448&1.883&0.672&41.22$\pm$5.44&0.03\\
         20069 (11)&150.420&1.974&0.863&29.82$\pm$9.86&0.12\\
         \hline
    \end{tabular}
    \label{tab:X-ray cluster}
\end{table*}

Then we compare our sample with optically selected cluster catalogues. There are three matches in \citet{Bellagamba_2011}.
In this work, the clusters are selected by weak lensing data.
Our cluster candidates 05, 06, and 07 match with clusters 17, 19, and 24, respectively, in their work. The information of these three clusters is listed in Table \ref{tab:optical cluster}. Since the weak lensing method detects clusters through dark matter spatial distribution rather than through galaxy overdensities, the difference is expected.

\begin{table}
    \centering
    \caption{Properties of weak lensing selected clusters taken from \citet{Bellagamba_2011}. In the first column, brackets are our own cluster IDs listed in Table \ref{tab:cluster sample}. The last column shows the distance between the \citet{Bellagamba_2011} and our definition of the cluster centre.}
    \begin{tabular}{ccccc}
         \hline
         ID & RA & DEC & ${\rm z}_{\rm spec}$ & Deviation \\
         &[degree]&[degree]&& [${\rm Mpc}$]\\
         \hline
         17 (05)&150.590&2.473&0.62&0.26\\
         19 (06)&150.443&1.883&0.64&0.10\\
         24 (07)&149.921&2.521&0.72&0.10\\
         \hline
    \end{tabular}
    \label{tab:optical cluster}
\end{table}

We also compared our cluster candidates with the COSMOS environment catalogue \citep{Darvish_2017}, where the authors use photometric data from \citet{Laigle_2016} to construct a density field. They identify the environment (cluster, filament, field) from the density field for each galaxy without giving clusters' information. We slice the galaxy redshifts as in this work, plot cluster galaxies in each slice and compare them with our cluster candidates. Our cluster candidates, 01, 02, 04, 05, 06, 07, 08, 10, 14, 15, and 17 match well with dense regions of "cluster" galaxy distribution. The other cluster candidates also contain "cluster" galaxies in the aperture, but their centres are offset from the densest regions.

In conclusion, 12 of our 17 cluster candidates have good matches with the clusters discovered in previous studies. In previous works, all 17 cluster candidates have a cluster match within 1 Mpc.

\section{Methods}\label{sec:method}
In Sect. \ref{sec:UVJ}, we describe how we classify recently-quenched galaxies, star-forming galaxies, and quiescent galaxies. In Sect. \ref{sec:fs}, we explain how we separate fast and slow quenching galaxies on the UVJ diagram. In Sect. \ref{sec:subtraction}, we explain how we remove the impact of possible projection effects.

\subsection{Photometric selection of RQGs}\label{sec:UVJ}
\subsubsection{Rest frame colour index estimation}\label{sec:colour estimation}
We use the EAZY code \citep{Brammer_2008} which fits SED by maximizing the likelihood between observed and template predicted SED. The templates of EAZY are set purely on stellar population synthesis models, which suit deep optical-NIR surveys. A non-negative matrix factorization (NMF) algorithm \citep{BR07} is adopted to reduce the number of templates while keeping their ability to reproduce observation. For input of NMF, the library of PEGASE models described in \citet{Grazian_2006} is used. This library includes $\sim$3000 models with ages from 1 Myr to 20 Gyr and several models for SFHs. These templates also consider dust extinction using the \citet{Calzetti_2000} law and include emission lines. The EAZY code fits SED using a linear combination of these templates. Through the SED fitting, rest-frame UVJ colours can also be derived. In this work, we use rest-frame U and V filter information from \citet{Johnson_2006} and rest-frame J band from 2MASS survey \citep{Skrutskie_2006}. We fix the redshifts of galaxies to the photometric redshifts estimated in the COSMOS2015 catalogue \citep{Laigle_2016}. As a result, EAZY provides interpolated colour index L$_X$ (X=U, V, J). Then, U$-$V and V$-$J colour indices are calculated from U, V, and J colour indices with the form:
\begin{equation}
{\rm X}-{\rm Y}=-2.5\times {\rm log} \frac{L_{\rm X}}{L_{\rm Y}}\qquad{\rm (X, Y = U, V, J)}
\end{equation}

\subsubsection{Rest-frame UVJ diagram}\label{sec:UVJ diagram}
\begin{figure}
	\includegraphics[width=\columnwidth]{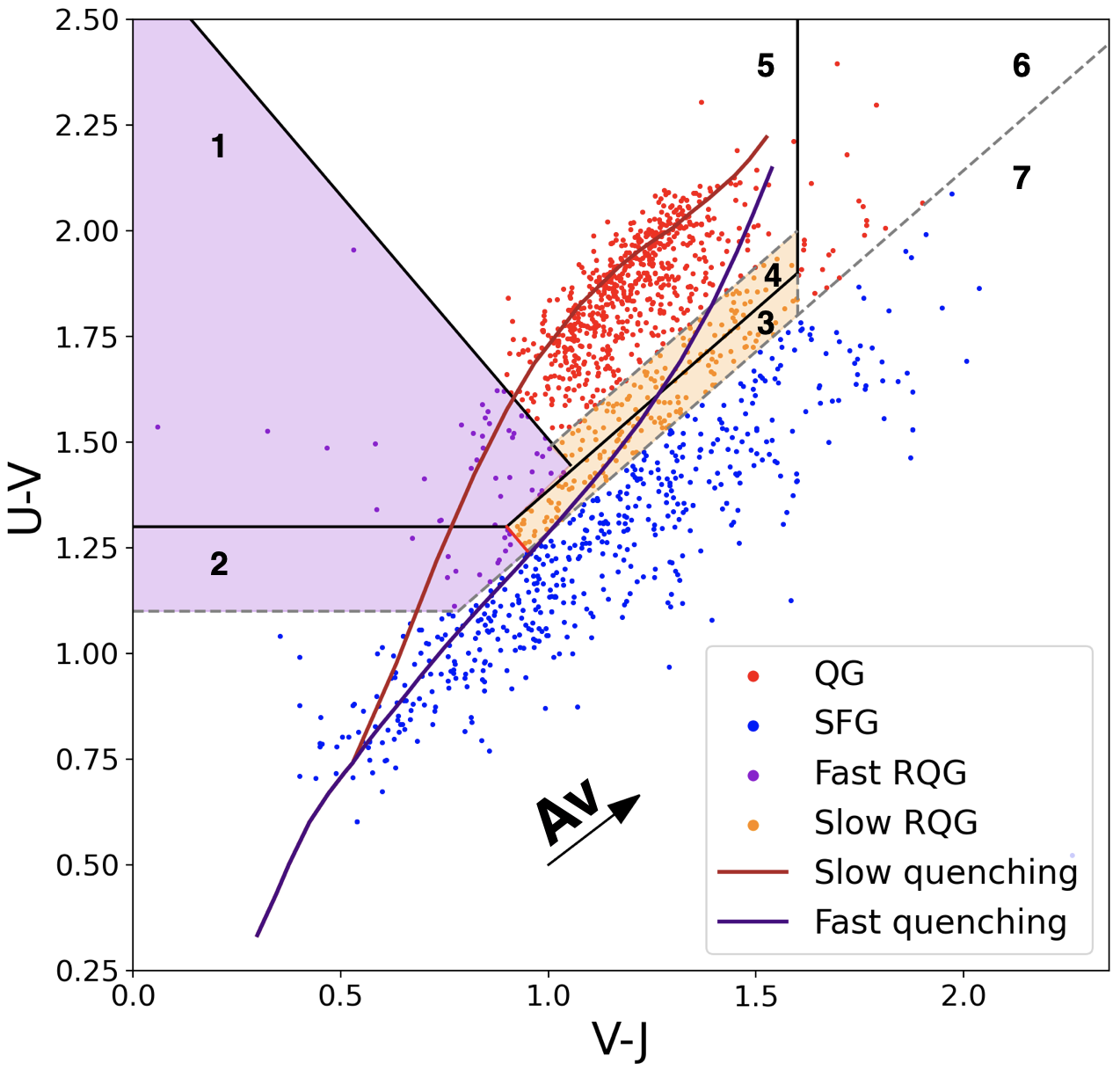}
    \caption{ Rest-frame UVJ diagram used to classify galaxy populations. The solid black lines are boundaries in \citet{Schreiber_2018}. Dashed lines are modified boundaries we use in this work (see more in Sect. \ref{sec:UVJ diagram}). Red dots are quiescent galaxies; blue dots are star-forming galaxies; purple and orange dots are both recently quenched galaxies. The colour-shaded regions are RQG regions, and they are divided into fast and slow quenching regions. The two regions are divided by black and red diagonal solid lines according to evolution tracks (red and purple lines) of two model galaxies with 1 and 0.1 Gyr quenching timescale, respectively (see Sect. \ref{sec:fs}). The arrow is the dust extinction vector corresponding to $\rm E(B-V)=0.1$. Numbers mark different regions. 1 and 2 are fast quenching RQG; 3 and 4 are slow quenching RQG; 6 and 7 are QG; 3 is RQG.}
    \label{fig:UVJ}
\end{figure}
The rest-frame UVJ diagram is a common tool to separate quiescent and star-forming galaxies (e.g. \citealt{Daddi_2004, Patel_2012, Arnouts_2013}). Quiescent galaxies are redder in U$-$V colour for a given V$-$J colour due to their strong Balmer break feature. Since RQGs are transitional galaxies between these two populations, a natural thought is that they should also reside in a region between these two populations on the rest-frame UVJ diagram.

The location of RQGs on the UVJ diagram (Fig. \ref{fig:UVJ}) has been discussed in many works, like \citet{Almaini_2017}, \citet{Schreiber_2018}, \citet{Belli_2019} and \citet{Wu_2020}. These works show the UVJ colours of RQGs selected by the Principal Component Analysis (PCA) method (based on supercolours derived from galaxy photometry that correlates with galaxy spectral features, mainly D4000 and absorption line), specific star formation rate (sSFR) obtained from SED fitting, Keck/MOSFIRE spectra, and LEGA-C spectra respectively. It is noted that the boundaries between different populations are arbitrary. Taking advantage of the existing comparisons, we set our own selection boundaries to include as many RQGs as possible while keeping a relatively low contamination level (see Sect. \ref{sec:confirmation}). 

In this paragraph, we briefly introduce how our boundaries are decided (Fig. \ref{fig:UVJ}). We modified the boundaries from \citet{Schreiber_2018} (solid black lines), which separates SFGs (region 2, 3, 6, and 7) and QGs (region 1, 4, and 5). In this work, dusty and non-dusty galaxies are divided by a diagonal line. We restrict it within the quiescent region to separate the post-starburst galaxies (region 1) and fully quenched QGs (region 4 and 5) as in \citet{Wild_2014} and \citet{Almaini_2017}.
However, some works \citep{Almaini_2017,Schreiber_2018,Wu_2020} identified RQGs outside the solid boundaries in both star-forming and quiescent regions by different methods (e.g. PCA method, SED fitting, spectral features). Therefore, we shift the original solid horizontal boundary to U-V=1.1 (grey dashed horizontal line) and shift the diagonal boundary by $\rm \pm\Delta(U-V)=0.1$ (grey dashed diagonal lines) to add regions 2, 3, and 4 into the original post-starburst region 1. The final RQG sample (regions 1, 2, 3, and 4) should fulfil all the following conditions:

\begin{equation} \label{eq:UVJ}
\left\{
\begin{array}{lr}
    U-V>1.1, & \\
    U-V>0.857(V-J)+0.429, & \\
    V-J<1.6, & \\
    U-V<0.857(V-J)+0.629, & \\
    U-V<-1.154(V-J)+2.661
\end{array}
\right.
\end{equation}

Fig. \ref{fig:UVJ} depicts all the boundaries used to classify these three galaxy populations.
The star-forming galaxies (blue points) lie in region 7 and show a sequence depending on dust extinction. The quiescent galaxies (red points) locate in regions 5 and 6, and the dusty quiescent galaxies locate in region 6 \citep{Almaini_2017}. The RQGs lie between the SFG and QG populations, i.e. in the middle of the evolutionary paths from SFGs to QGs, as shown by the solid curves (see Sect. \ref{sec:fs}). The RQG region (1, 2, 3, and 4) is marked by purple and orange shades. There is a concern that these arbitrary boundaries may lead to some uncertain results. However, we shift the boundaries between QG and RQG by 0.1 magnitudes and find that it does not affect our qualitative results.

\subsection{Fast and slow quenching galaxy separation}\label{sec:fs}
The quenching timescale is an important factor in understanding its physical mechanisms. To put a strict constraint on the quenching timescale, it is necessary to obtain high resolution, high SNR rest-frame optical spectra to fit with the galaxy model. However, the number of our sample galaxies with spectra is limited. To give an idea of how the quenching timescale depends on different parameters, we take a photometric approach and apply model colour tracks to separate galaxies with different quenching timescales.

\begin{figure}
	\includegraphics[width=\columnwidth]{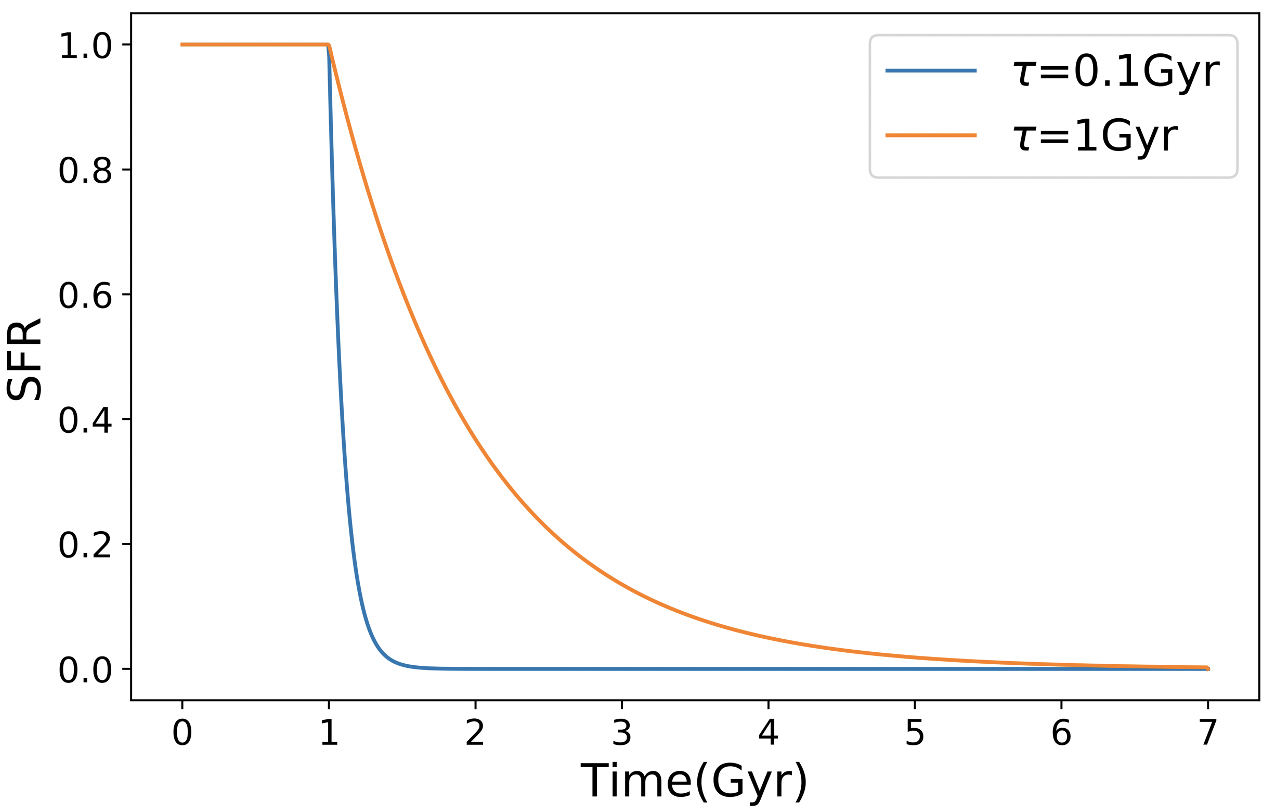}
    \caption{SFH models with a constant star formation period followed by exponentially declining quenching period with $\tau=0.1$ and 1 Gyr.}
    \label{fig:SFH}
\end{figure}

The SFH model we adopt has a constant star formation period followed by an exponentially declining star formation quenching phase, as illustrated in Fig. \ref{fig:SFH}. The star formation rate (SFR) evolution with age is set by Eq \ref{eq: SFH}.
\begin{equation}
{\rm SFR}(t)=\left\{
             \begin{array}{lr}
             1, & t < 1{ Gyr} \\
             exp (-\frac{t-1{ Gyr}}{\tau}), & t > 1{ Gyr}\\
             \end{array}
\right.
\label{eq: SFH}
\end{equation}
where $\tau$ is the quenching timescale. 

To understand how galaxies move in the rest-frame UVJ space, we use CIGALE code \citep{Boquien_2019} to make galaxy models and estimate their rest-frame colours. In the estimation of rest-frame UVJ colours, we use exactly the same filter as Sect. \ref{sec:colour estimation}. In this work, we assume the galaxies have \citet{Chabrier_2003} initial mass function, solar metallicity, and \citet{Calzetti_2000} dust attenuation curve with E(B$-$V)=0.15. Since the RQGs have some remaining star formation; there should be a certain amount of dust component. And we arbitrarily choose the value of E(B$-$V)=0.15 for galaxies in the RQG phase. More details of dust attenuation's effects on evolutionary paths are discussed in Appendix \ref{app:model track}. For a fast quenching galaxy, we assume a quenching timescale of 0.1 Gyr, and for a slow quenching galaxy, we assume a quenching timescale of 1 Gyr. In Appendix \ref{app:model track}, we discuss how different parameters affect evolutionary pathways in the UVJ space. The evolutionary pathways of the two model galaxies are plotted in Fig. \ref{fig:UVJ}. We can see there is a clear difference between the fast and the slow quenching tracks. To separate the fast (purple) and slow (orange) quenching regions, we adopt two lines. For regions 1 and 4, we adopt the solid diagonal line from \citet{Schreiber_2018}. For regions 2 and 3, we draw a solid red line with the same slope as the previous one where the slow quenching track enters the RQG region.

CIGALE also allows us to estimate the time when the model galaxies stay in the RQG selection area, and we define this time as the visibility timescale. When the quenching timescale is 0.1, 0.3, 1, 2 Gyr, the visibility timescale is 0.43, 0.58, 1.5, 2.6 Gyr respectively. Generally, galaxies with a longer quenching timescale have a longer visibility timescale. This trend will help us understand our results better. However, since the galaxy models here are simplified, the value of the visibility timescale will not be used for quantitative analysis.

\subsection{Field subtraction}\label{sec:subtraction}

In order to select galaxy cluster candidates we have used photometric redshifts adopting redshift intervals (full width) of $\Delta z = 0.1$ at $0.5<z<1$. However, the intrinsic scatter of a cluster in redshift space is much smaller than $\Delta z = 0.1$. Therefore, there should be many galaxies that are not physically associated with our cluster candidates in each redshift slice. Thus, we need to apply a statistical field subtraction to correct for such a projection effect. To achieve that, we first define field regions following the same method applied during the cluster candidate search. We use aperture density maps for each redshift slice and define the field area as the regions with densities lower than the $1\sigma$ threshold, which corresponds to the regions outside the cyan contours in Fig. \ref{fig:density map}. Information of the field samples in each redshift slice is shown in Table \ref{tab:field sample}.

\begin{table*}
\caption{Information of the field sample in each redshift slice. The comoving area and the number of each population are given for each slice. Fast RQG and slow RQG are recently quenched galaxies located in fast (purple) and slow (orange) quenching regions on the UVJ diagram (Fig. \ref{fig:UVJ}).}
\setlength{\tabcolsep}{1.7mm}{
\begin{tabular}{cccccccc}
\hline
\\[-3mm]
Redshift bin & Field area [${\rm Mpc}^2$] & log (Field density) [$/{\rm Mpc}^2$] & N(SF) & N(RQG)& N(QG) & N(fast RQG) & N(slow RQG)\\
\\[-3mm]
\hline
\\[-3.4mm]
0.5-0.6 & 845 & 0.36 & 971 & 339 & 628 & 71 & 268\\
0.55-0.65 & 889 & 0.34 & 977 & 199 & 751 & 60 & 139\\
0.6-0.7 & 972 & 0.49 & 1425 & 477 & 1099 & 114 & 363\\
0.65-0.75 & 1003 & 0.49 & 1609 & 432 & 1047 & 137 & 295\\
0.8-0.9 & 1134 & 0.55 & 1980 & 757 & 1286 & 281 & 476\\
0.85-0.95 & 1172 & 0.53 & 2183 & 668 & 1117 & 256 & 412\\
0.9-1.0 & 1228 & 0.57 & 2568 & 675 & 1284 & 293 & 382\\
\\[-3.5mm]
\hline\\
\end{tabular}}
\label{tab:field sample}
\end{table*}

Before performing the field subtraction, we divide our sample of cluster and field galaxies into several stellar mass bins, which will be used to carry out our environmental analysis in Sect. \ref{sec:result}. Finally, we compute the density of each population (SFG, RQG, QG) within each stellar mass bin and subtract the field contribution from each population using the following equation:

\begin{equation}
\begin{split}
N_{\rm subtracted}=N-\Sigma_k\times N_{\rm cluster}\times 2.8{\rm Mpc}\times2.8{\rm Mpc} \\
(k=SFG, RQG, QG),    
\end{split}
\end{equation}

where $N_{\rm subtracted}$ is the number of galaxies after the subtraction, N is the number of galaxies before the subtraction, $\Sigma_k$ is the field density of each population, and  $N_{\rm cluster}$ is the total number of the cluster candidates in this redshift bin. While studying the environmental dependence, galaxies will also be binned in local density ($\Sigma_5$) during our analysis. Thus, we need to subtract possible field galaxies from each local density bin too. After computing the field density of each population, we then subtract it by applying:
\begin{equation}
\begin{split}
N_{\rm subtracted}=N-\sum_i(\frac{\Sigma_{\rm field}}{\Sigma_5}\times\frac{\Sigma_k}{\Sigma_{SFG}+\Sigma_{RQG}+\Sigma_{QG}})\\
(k=SFG, RQG,QG),
\end{split}
\end{equation}
where $\Sigma_k$ is the field density of each population and $\Sigma_{field}$ is the field density. The first item in the bracket is the probability of a cluster candidate member galaxy being a field galaxy; the second item is its probability of being an SFG, an RQG, or a QG. After calculating this probability for each cluster candidate member, we then sum up this value for all galaxies in the local density bin. 

\section{Results}\label{sec:result}
In this section, we present the results of our analysis in our selected cluster candidate sample (Table \ref{tab:cluster sample}). The average field densities are displayed in Table \ref{tab:field sample}. 
All errors in this section are propagated based on the Poisson statistics for both the original data and the field subtraction using a standard method. The information of all bins we analyse in this section is listed in the tables in Appendix \ref{app:bin information}.

\subsection{Quenching efficiency}\label{sec:qe}
Galaxy quenching drives star-forming galaxies to evolve into recently-quenched galaxies. Thus, the ratio of these two populations can represent the efficiency of the quenching process. Hence, we define the quenching efficiency as:
\begin{equation}
 Q.E. = \frac{N_{RQG}}{N_{SFG}}
\end{equation}
\begin{figure*}
	\includegraphics[width=2\columnwidth]{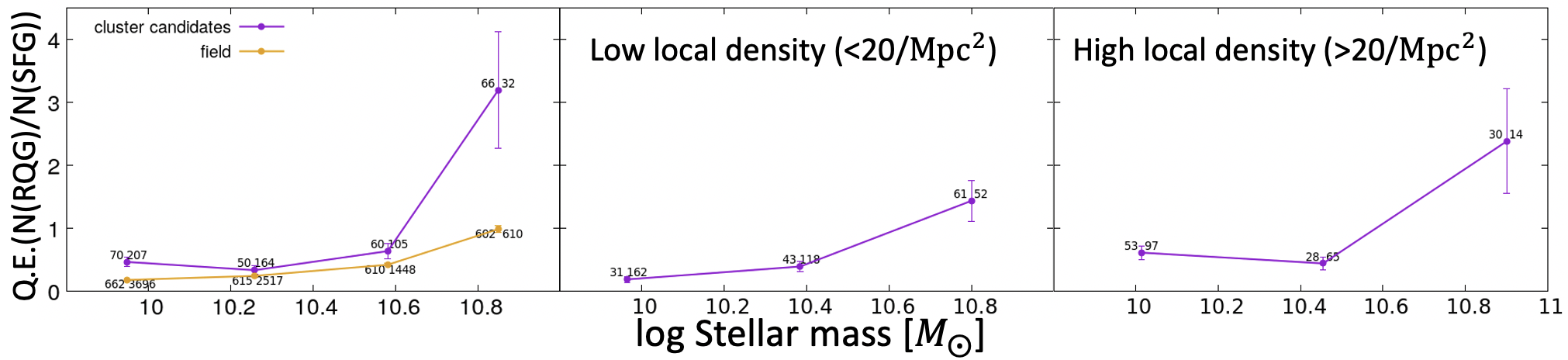}
    \caption{This figure displays quenching efficiency's mass dependence. The left panel shows dependence in all environments in our sample cluster candidates. The middle and right panels show dependence in two subsamples divided by local density. The total numbers of galaxies in the two subsamples are the same. The numbers to the left of each data point are the numbers of RQGs in this mass bin, while the numbers to the right are those of SFGs. The total numbers of SFGs or RQGs in the middle and the right panels are the same as the numbers in the cluster candidates in the left panel. Errors are propagated based on the Poisson statistics error of each population in each bin. 
    }
    \label{fig:qemass}
\end{figure*}

Figure \ref{fig:qemass} presents how Q.E. depends on stellar mass. The left panel also shows a comparison between the cluster candidates and the general field. Quenching efficiency shows a strong dependence on the stellar mass, especially in the most massive bin. The quenching efficiency is always higher in the cluster candidates, showing that the cluster candidate environment is enhancing it. The gap is much more prominent in the most massive bin, indicating the mass quenching is also stronger in the cluster candidates. To see how the local galaxy density affects the mass-dependent quenching, we divide our sample into high- and low-density subsamples at $10^{1.3}/Mpc^2$ (i.e. 20$/Mpc^2$), each of which contains one half of the galaxy sample (middle and right panels in Fig. \ref{fig:qemass}). It turns out that the local environment does not have a substantial impact on mass-dependent quenching. And there is a significant rise in the quenching efficiency with stellar mass for both sparse and dense local environments.

\begin{figure*}
	\includegraphics[width=2\columnwidth]{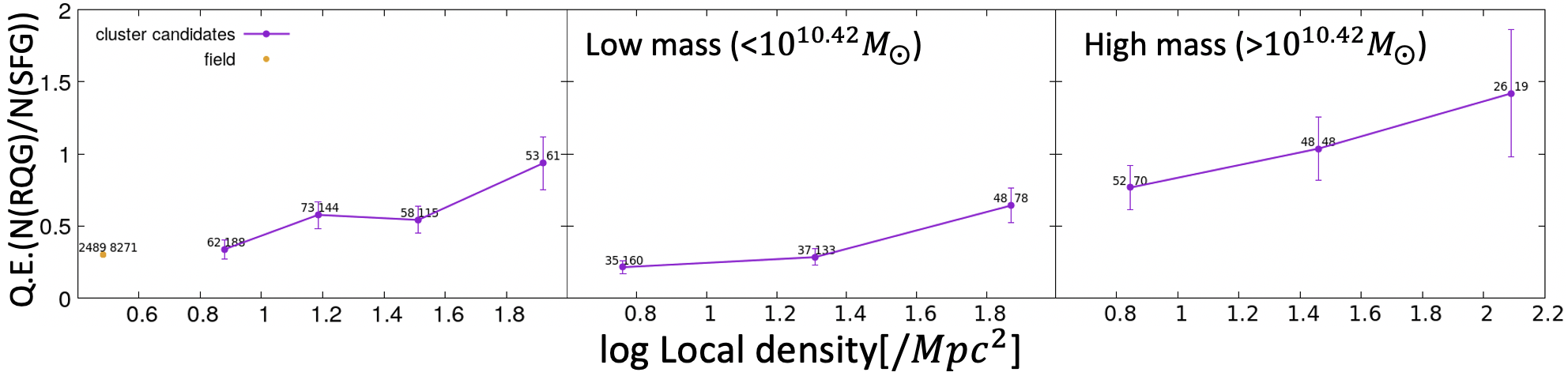}
    \caption{This figure displays the quenching efficiency's environmental dependence. The left panel shows the dependence of all masses. The middle and right panels show the dependence in two subsamples divided by stellar mass. The total numbers of galaxies in the two subsamples are the same. The numbers to the left of each data point are numbers of RQGs in this mass bin, while the numbers to the right are numbers of SFGs.  The total numbers of SFGs or RQGs in the middle and the right panels are the same as the numbers in the cluster candidates in the left panel. Errors are propagated based on the Poisson statistics for each population in each bin. 
    }
    \label{fig:qesigma}
\end{figure*}

In the left panel of Fig. \ref{fig:qesigma}, the quenching efficiency in the cluster candidates shows a dependence on the local density, which supports the scenario of environmental quenching. In the field, the quenching efficiency is slightly lower than the lowest local density bin in the cluster candidates, indicating the outskirts of the cluster candidates have a similar enhancement in galaxy quenching as the field. The sample is divided into a low mass and a high mass subsamples at $M_*=10^{10.42}M_{\odot}$ for analysis. In both low mass and high mass subsamples, quenching efficiency increases with local density. However, considering the errorbars, we can only say that the environmental dependence in low mass systems (the middle panel of Fig. \ref{fig:qesigma}) is significant. If we compare the quenching efficiency in the two subsamples, high mass galaxies have higher quenching efficiency, which again confirms the presence of mass-dependent quenching.

\subsection{Quenching stage}\label{sec:qs}
When the quenching process is finished, RQGs will become QGs. Therefore, we define {\it quenching stage} as the number ratio of these two populations:
\begin{equation}
Q.S. = \frac{N_{RQG}}{N_{QG}}
\end{equation}
where the Q.S. value indicates how much the quenching process has proceeded for a given galaxy sample. When the Q.S. value is high, many galaxies are in the RQG phase and undergoing quenching, indicating a relatively early stage of quenching. On the other hand, if the Q.S. value is low, it means that most of the galaxies are already quenched and have become old QGs.
\begin{figure*}
	\includegraphics[width=2\columnwidth]{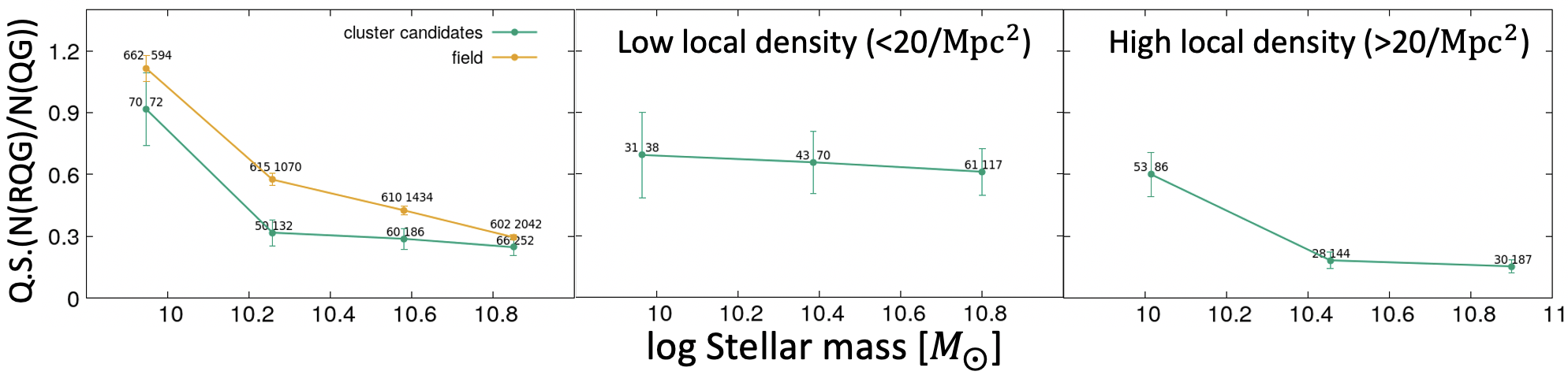}
    \caption{This figure displays the quenching stage's mass dependence. The left panel shows dependence in all environments in our sample of cluster candidates. The middle and right panels show the dependence in two subsamples divided by local density (same bin division as in Fig. \ref{fig:qemass}). The numbers to the left of each data point are the numbers of RQGs in this mass bin, while the numbers to the right are those of QGs. The total number of QG or RQG in the middle and the right panels is the same as the number of cluster candidates in the left panel. Errors are propagated based on the Poisson statistics for each population in each bin. 
    }
    \label{fig:qsmass}
\end{figure*}

As it can be seen in the left-hand panel of Fig. \ref{fig:qsmass}, the quenching stage decreases with the stellar mass, especially for the least massive bin. The quenching stage in the field is higher than in the cluster candidates, showing they are in an earlier phase of quenching. Lowest mass galaxies are in an earlier stage of quenching, which is consistent with the downsizing scenario in which massive galaxies form and quench earlier than low mass ones. 

We then try to reveal how the local environment affects this relation (Fig. \ref{fig:qsmass}, centre and right diagrams). The full sample was divided into a high and a low local density subsample using $\Sigma_5$ at $20/{Mpc}^2$ as a threshold, and ensuring that both subsamples contain half of all galaxies. The trend between the quenching stage and stellar mass is only significant for the denser local environment. 
\begin{figure*}
	\includegraphics[width=2\columnwidth]{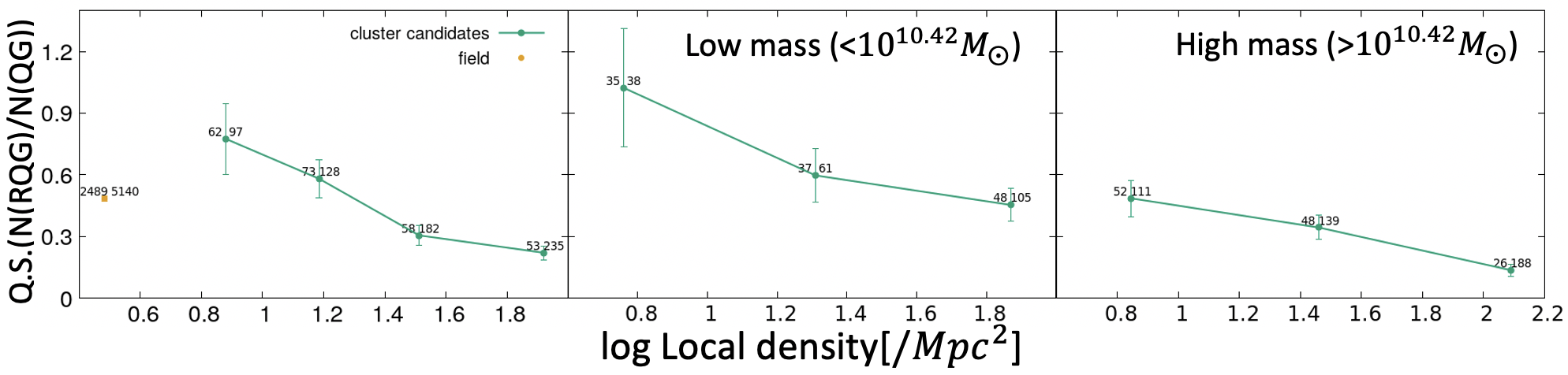}
    \caption{This figure displays the quenching stage's local density dependence. The left panel shows the dependence for all environments. The middle and right panels show dependence in different mass bins (same bin division as in Fig. \ref{fig:qesigma}). The numbers to the left of each data point are numbers of RQGs in this density bin, while the numbers to the right are numbers of QGs. The total number of QG or RQG in the middle and the right panels is the same as the number of cluster candidates in the left panel. Errors are propagated based on the Poisson statistics for each population in each bin. 
    }
    \label{fig:qssigma}
\end{figure*}
As in Fig. \ref{fig:qssigma}, the quenching stage clearly decreases with local density. If we separate massive and less massive subsamples at $M_*=10^{10.42}M_{\odot}$ (Fig. \ref{fig:qssigma}), the trends are still seen in both subsamples, but are less significant due to decreased sample sizes. The quenching stage in the field is lower than the least dense cluster bin. However, the current field sample size is large, and the errorbar will increase if it has a similar size as the cluster. Therefore, this result is not significant and we do not further discuss it here. We also notice that mass dependence shows different properties in two local density bins, and environment dependence shows different values in two mass bins. Hence, the mass dependence of quenching also has a dependence on the environment, and the environment dependence of quenching also has a dependence on stellar mass.

\subsection{Quenching timescale}\label{sec:timescale}
\subsubsection{Mass, environmental, and redshift dependence}\label{sec:timescale dependence}
\begin{figure}
	\includegraphics[width=\columnwidth]{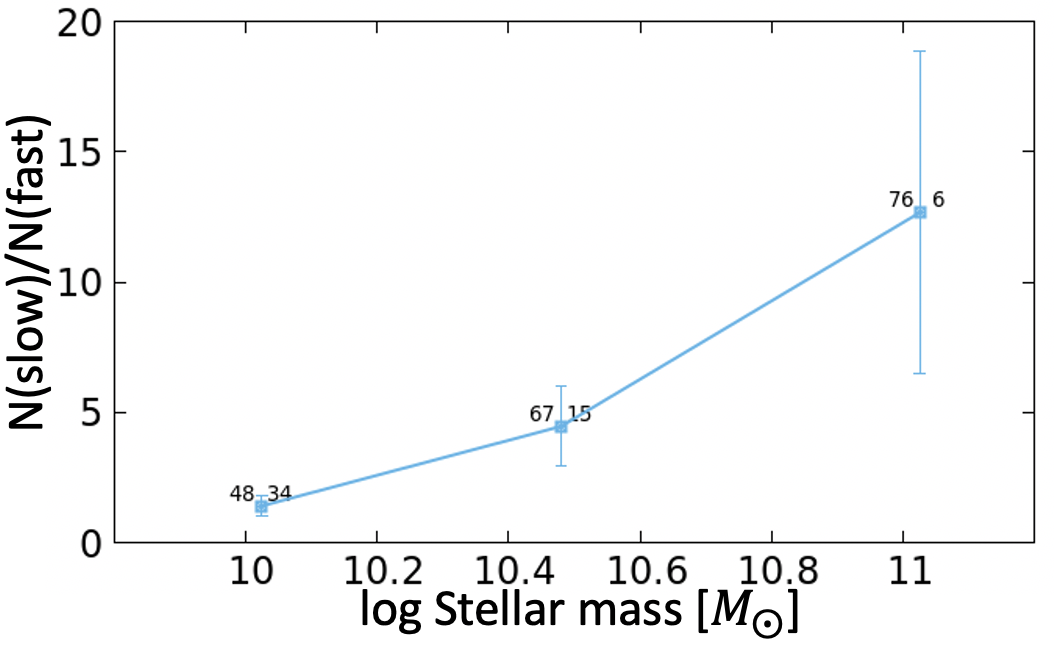}
    \caption{This figure shows quenching timescale increases with stellar mass. The numbers to the left and right of each data point respectively show the number of slow and fast quenching galaxies for each mass bin. Errors are derived based on the Poisson statistics for each population in each bin. 
    }
    \label{fig:fsmass}
\end{figure}
\begin{figure}
	\includegraphics[width=\columnwidth]{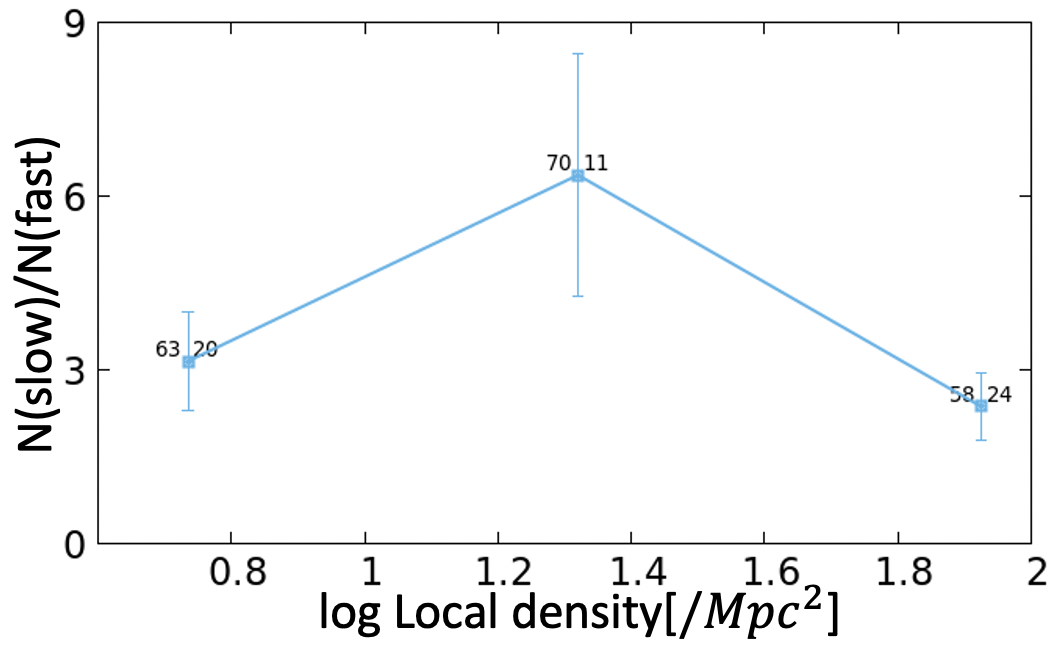}
    \caption{The quenching timescale does not have a clear relation with local density. The numbers to the left and right of each data point respectively represent the number of slow and fast quenching galaxies for each mass bin. Errors are derived based on the Poisson statistics error of each population in each bin.
    }
    \label{fig:fssigma}
\end{figure}
\begin{figure}
	\includegraphics[width=\columnwidth]{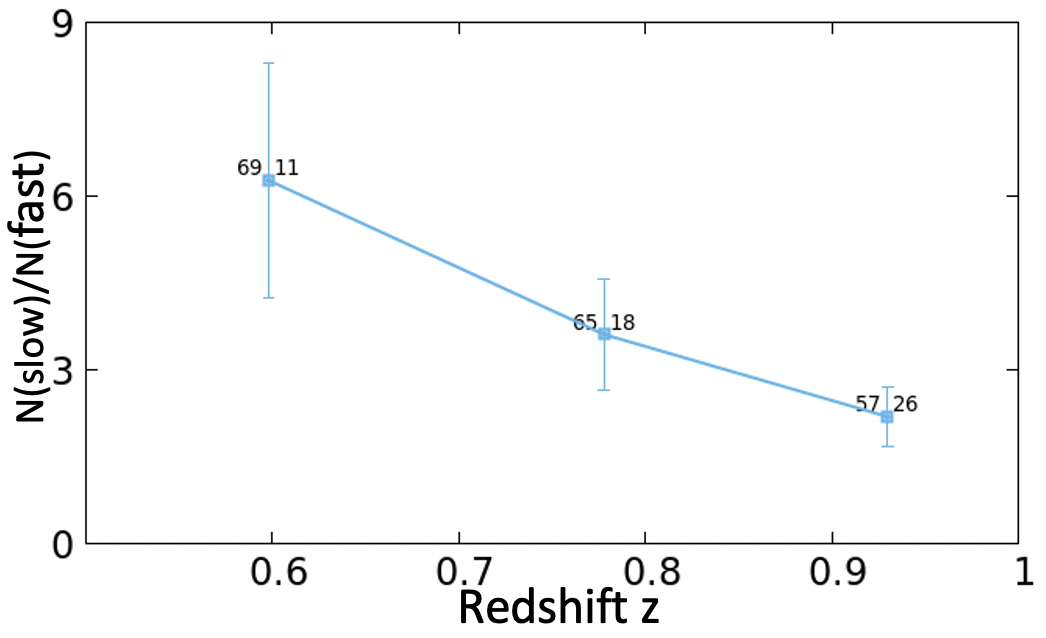}
    \caption{This figure shows quenching timescale decreases with redshift. The numbers to the left and right of each data point respectively represent the number of slow and fast quenching galaxies for each mass bin. Errors are derived based on the Poisson statistics error of each population in each bin.
    }
    \label{fig:fsz}
\end{figure}
As we have emphasised, the quenching timescale is an important factor to discriminate among different physical mechanisms of quenching. Although we separate fast and slow quenching galaxies only based on the rest-frame UVJ diagram, we can still use the number ratio of these two populations to illustrate the qualitative dependence of quenching timescale on mass and environment.

In Fig. \ref{fig:fsmass}, we show that the number ratio of slow quenching galaxies over fast quenching galaxies increases with the stellar mass, which means low mass galaxies have shorter quenching timescales compared to those of massive galaxies. This suggests that the quenching of galaxies with different stellar mass is dominated by different physical mechanisms.

In Fig. \ref{fig:fssigma}, there is a possible dependence of the quenching timescale on the local density. The ratio seems to have a peak in the intermediate density bin with large uncertainty. However, we cannot discuss if it is related to any physical trend since our sample size is still limited. 

In Fig. \ref{fig:fsz}, we show there is a clear trend that fast quenching is more important at higher redshifts. This suggests that the main quenching mechanism changes with redshift, and the short timescale quenching becomes more important at higher redshifts, which is consistent with the results in \citet{Cucciati_2012} and \citet{Rowlands_2018}. 

\subsubsection{Two specific cases}\label{sec:cases}
We here highlight two specified cases in quenching which show the largest difference in the slow and fast quenching ratio, as they probably correspond most directly to the mass and environment-dependent quenching, respectively.
\begin{table}
\centering
\caption{N(slow)/N(fast) values and their errors in the two specified cases. The high mass and low density case may be more affected by mass-dependent quenching, and the low mass high density case may be more affected by environmental quenching. Details are discussed in Sect. \ref{sec:cases}.}
\begin{tabular}{ccc}
\hline
\\[-3mm]
Description & N(slow)/N(fast)\\
\\[-3mm]
\hline
\\[-3.3mm]
high mass + low local density & 6.500 $\pm$ 2.264 \\
low mass + high local density & 1.308 $\pm$ 0.343 \\
\\[-3.3mm]
\hline\\
\end{tabular}
\label{tab:slowfast}
\end{table}

In Table \ref{tab:slowfast}, the first case is the quenching of massive galaxies in a low density environment. High mass refers to the higher 50\% of mass (i.e. $>10^{10.42}M_{\odot}$). Low local density refers to the lower 50\% of local density (i.e.$<10^{1.3}/Mpc^2$). This case should be mainly affected by mass-dependent quenching. The N(slow)/N(fast) ratio has a high value, giving a hint that the mechanism working in this case has a long timescale. {This is likely to be, for example, AGN feedback, considering AGNs are exclusively hosted by massive galaxies, and the AGN feedback can be a slow quenching process with a timescale of a few Gyr \citep{Hirschmann_2017}. }

The second case is the quenching of low mass galaxies in a dense environment, which is expected to be more affected by the environment. The N(slow)/N(fast) ratio is much lower compared to the first case, showing the impact of a short timescale mechanism. It is likely to be, for example, ram pressure stripping, which strips gas from low mass galaxies easier due to their shallower gravitational wells. Moreover, ram pressure stripping can only occur in dense cluster cores on a relatively short timescale \citep{Steinhauser_2016}. However, we should notice that short timescale AGN feedback quenchings also exist; they can play a role in this case as well.

We note that we cannot reject other possibilities since the information we are able to obtain with photometric data alone is still limited.

\subsection{Visibility timescale}\label{sec:visibility}
Apart from the reasons we analysed in the previous sections, we should notice that the visibility timescale of galaxies can also play a role in our results. Visibility timescale is a galaxy's timescale of staying in the RQG region on the UVJ diagram. If the visibility timescale also depends on stellar mass or local density, it would degenerate with the quenching efficiency and quenching stage. In Sect. \ref{sec:fs}, we show that galaxies with longer quenching timescale generally spend longer time in the RQG region on the UVJ diagram. However, we should note that with photometric data solely, we cannot constrain visibility time with a small enough uncertainty to completely break the degeneracy between mass and environment dependence of visibility time and quenching properties. Hence we only separate the fast quenching (short visibility time) and the slow quenching (long visibility time) bins and estimate their contributions to know how the visibility timescale affects our results.

\begin{figure*}
	\includegraphics[width=2\columnwidth]{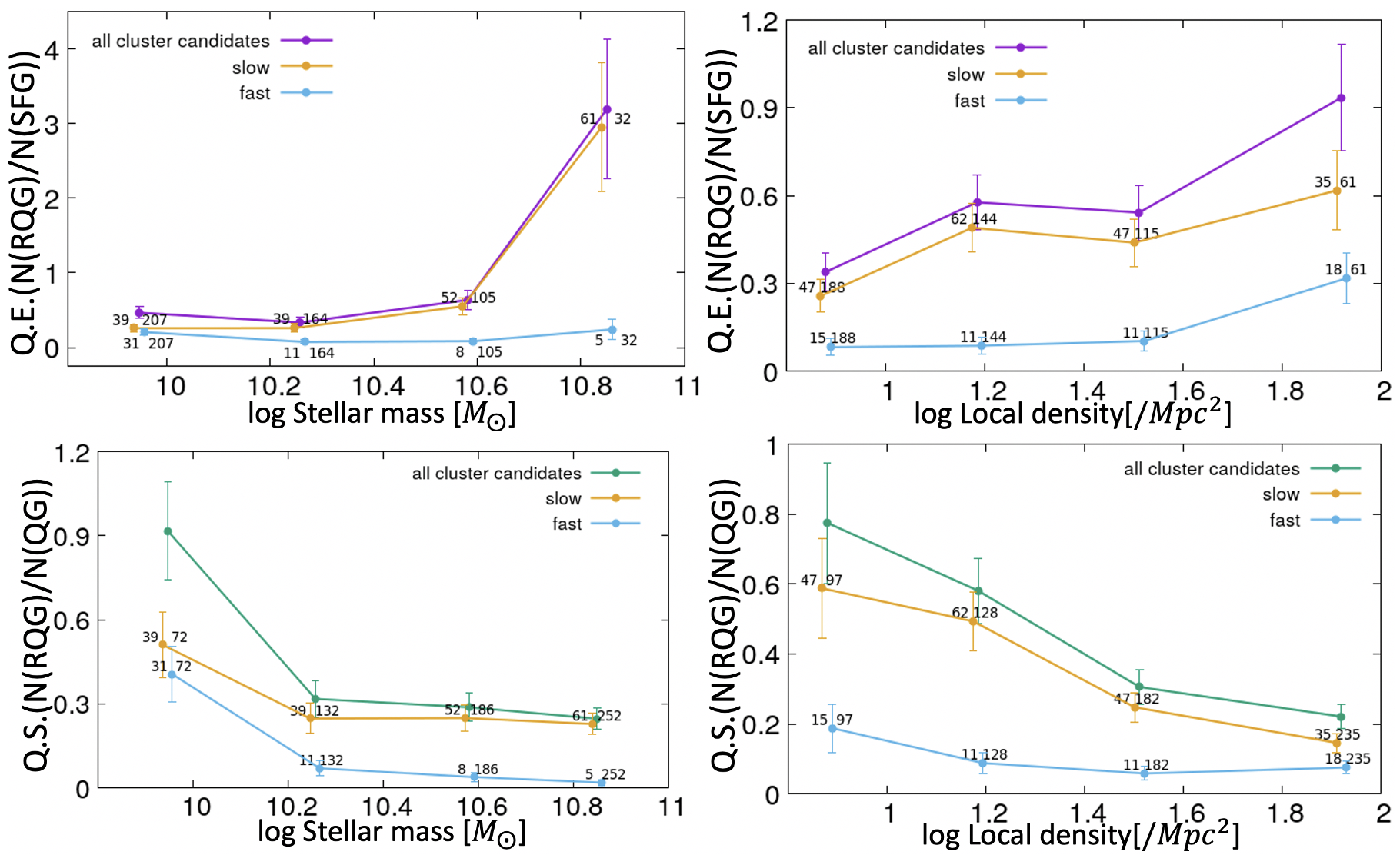}
    \caption{In this figure, we show results of quenching efficiency and quenching stage in fast and slow quenching galaxies separately. Slow quenching galaxies (same as long visibility timescale galaxies) are marked with orange points, while fast quenching galaxies (same as short visibility timescale galaxies) are marked with light blue points.}
    \label{fig:fast slow separated}
\end{figure*}

In Fig. \ref{fig:fast slow separated}, we show how the fast quenching (short visibility time) galaxies and the slow quenching (long visibility time) galaxies contribute to the mass and the environment dependence of $Q.E.$ and $Q.S.$. In the mass dependence of the quenching efficiency panel (top left panel), we can see that the long visibility time galaxies have a strong mass dependence while the short visibility time galaxies do not. It shows that the visibility time does affect the quenching efficiency; the increasing trend of quenching efficiency is partly due to the mass dependence of visibility time, and the massive galaxies tend to stay in the RQG region for longer. In the local density dependence of the quenching efficiency panel (top right panel), both fast and slow quenching galaxies show an increasing trend, indicating that the visibility time does not have a strong impact on the environment dependence of Q.E.. This is consistent with Sect. \ref{sec:timescale}, where we show that the quenching timescale has a dependence on the stellar mass but not on the local density. 
In the quenching stage panels (bottom two panels), both fast and slow quenching galaxies show a decreasing trend (similar to the overall trend), showing that the visibility time does not strongly degenerate with the mass or environment dependence of the quenching stage.

For results in Sect. \ref{sec:timescale dependence}, degeneracy between quenching timescale and visibility timescale exists as well. However, there is no proper way to break this degeneracy with only galaxy photometry. We should keep this degeneracy in mind when interpreting the physical scenario of quenching.

In conclusion, while analysing the mass dependence of quenching efficiency, we should be careful of the influence of RQG visibility time. We leave this issue to our future work. When we have the spectra of RQGs, we can disentangle the degeneracy between the quenching properties and the visibility time.

\subsection{RQG comoving number density}\label{sec:number density}
As mentioned in Sect. \ref{sec:subtraction}, the numbers of all RQGs, fast quenching RQGs, and slow quenching RQGs all show relations with redshift. To better understand this dependence, we calculated the comoving number densities in each redshift bin. Note that in this section, we use not only galaxies in our 17 cluster candidates but all galaxies in the COSMOS field with $M_*>10^{10}M_{\odot}$.

\begin{figure}
	\includegraphics[width=\columnwidth]{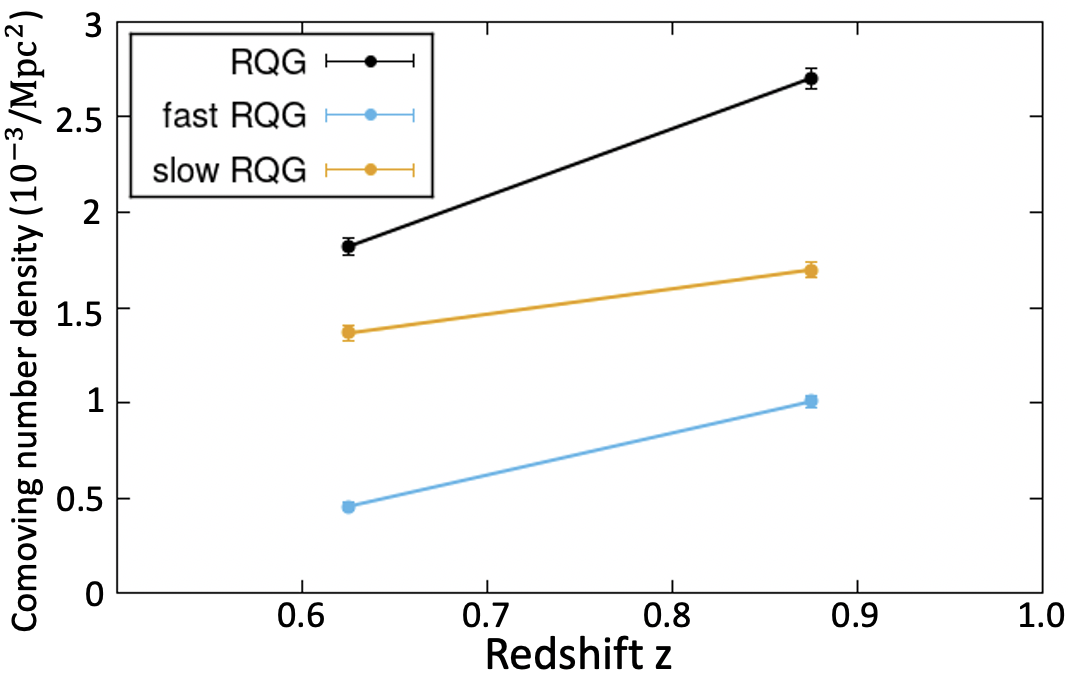}
    \caption{Comoving number densities' evolution of all RQGs, slow quenching RQGs and fast quenching RQGs. To increase the statistics, we bin all galaxies into two bins, $0.5<z<0.75$ or low redshift bin and $0.75<z<1.0$ or high redshift bin, consisting of 1468 and 2419 RQGs respectively}.
    \label{fig:density}
\end{figure}

In Fig. \ref{fig:density}, we present the number densities of fast quenching galaxies, slow quenching RQGs, and total RQG sample in different redshift bins (including both field and cluster candidate regions). 
We bin all galaxies into two groups to increase the statistics. The low-z bin compass $0.5<z<0.75$ while the high-z bin refers to $0.75<z<1.0$. Both slow and fast quenching RQGs show an increasing trend with redshift, and the fast quenching RQGs show a steeper slope. The increasing trend of fast quenching RQGs is consistent with \citet{Belli_2019} and \citet{Wild_2016} who revealed that the number density of post-starburst galaxies (corresponding to fast quenching RQGs in this work) increases with redshift. Comparing the slope of slow and fast quenching galaxies, we can tell that, similar to the trend in cluster candidate regions, fast quenching is more important at higher redshifts (see Sect. \ref{sec:timescale dependence}). 

\section{Discussions}\label{sec:discussion}
In Sect. \ref{sec:confirmation}, we discuss the contamination and completeness of our photometric RQG selection method. In Sect. \ref{sec:mechanism}, we discuss the physical implications of our results.
\subsection{Spectroscopic confirmation of our RQG selection} \label{sec:confirmation}
Our current result is based on photometric data. However, the UVJ selection is not fully confirmed by galaxy spectra, while the only way to precisely determine a galaxy's RQG properties is through its spectral features. The main limitation of RQG research is that it is a rare population, making it expensive to obtain a large sample of RQG spectra. Fortunately, the COSMOS field already has deep spectroscopic data available for many galaxies, enabling us to match our targets with the existing spectra.

We use spectroscopic data from DEIMOS 10K spectroscopic survey catalogue \citep{Hasinger_2018} and LEGA-C survey data release 3 \citep{Vanderwel_2021}. The DEIMOS 10K survey obtained the spectra of some of the COSMOS2015 catalogue galaxies. Depending on the galaxy, the spectral resolution can be either R$\sim2000$ or $\sim2700$, which are both large enough to measure the equivalent width of absorption lines. LEGA-C survey obtained galaxy spectra with R=2500. The parent sample LEGA-C adopts for observation is the UltraVISTA catalogue presented in \citet{Muzzin_2013}. 

The spectral features characteristic of RQGs are strong Balmer absorption lines \citep{Couch_1987}. We measure the galaxies' H$\delta$ absorption line equivalent width (EW), representing the Balmer absorption. For all the galaxies within our sample, galaxies with other strong Balmer absorption lines (e.g. H$\gamma$, H$\beta$ absorption line) have strong H$\delta$ absorption lines as well. Another feature of RQGs is that their star formation activities are at a low level, so we use [OII] emission line strength to examine the existence/absence of star formation activity in the galaxy in this study. We define a galaxy with EW(H$\delta$)<-4\AA\ and EW([OII])<5\AA\ (negative value represents absorption) as an RQG galaxy in this section.

The equivalent width of each line feature is calculated using
\begin{equation}
{\rm EW}=\int_{\lambda_1}^{\lambda_2}(1-\frac{F_{I,\lambda}}{F_{C,\lambda}})d\lambda,
\end{equation}
together with Lick spectral indices \citep{Salaris_2005} and the index of the [OII] line from \citet{Balogh_1999}. $F_{I,\lambda}$ is the index flux, which is obtained from the index band. $F_{C, \lambda}$ is the continuum flux, which we assume to be constant over the index band. The continuum flux is calculated from the average flux over blue and red continuum bands. We do not remove the infilling emission in the calculation of EW($H\delta$). All the spectral index information is listed in Table \ref{tab:line index}.

\begin{table*}
\caption{Spectral indices of RQG feature lines used in this study.}
\centering
\begin{tabular}[width=\columnwidth]{cccc}
\hline
\\[-3mm]
Name & Index band(\AA) & Blue continuum(\AA) & Red continuum(\AA)\\
\\[-3mm]
\hline
\\[-3.3mm]
H$\delta$ & 4083.50 - 4122.25 & 4041.60 - 4079.75 & 4128.50 - 4161.00\\
${\rm \lbrack OII\rbrack}$ & 3713.00 - 3741.00 & 3653.00 - 3713.00 & 3741.00 - 3801.00\\
\\[-3.3mm]
\hline\\
\end{tabular}
\label{tab:line index}
\end{table*}

For DEIMOS 10K observation, apart from matching our RQG sample with their spectroscopic catalogue, we also check spectra they marked with the H$\delta$ line feature. We have 105 matches in DEIMOS 10K catalogue. For the LEGA-C survey, we only match our RQG sample with their catalogue. We match the target positions within 1", and we have 23 matched RQGs. However, we noticed that there are spectra in both surveys that cannot provide enough S/N for accurate absorption line EW measurement due to various reasons. Especially the DEIMOS 10K spectra mostly have exposure times from 1 to 3 hours which is not long enough for less luminous galaxies. To avoid contamination from bad quality spectra, we select galaxies with EW errors of all four measured spectral lines $\sigma_i<0.8$\AA\ (i=H$\delta$, [OII]). This error cut leaves us with 40 high-quality spectra (28 from DEIMOS 10K and 12 from LEGA-C). Another thing we should notice is that the DEIMOS spectrograph has two chips that will produce an artificial break in the spectrum around 7600\AA\ . To avoid this effect, we checked all good quality spectra and remove those with this break in either the continuum or the index band of any of the four lines. There are RQG spectra without H$\delta$ coverage in both DEIMOS 10K and LEGA-C, we remove these galaxies as well. In total, we are left with 27 UVJ-selected RQGs with good enough spectra for spectroscopic confirmation. Their information is listed in Appendix \ref{app:spectra information}.

\begin{figure}
	\includegraphics[width=\columnwidth]{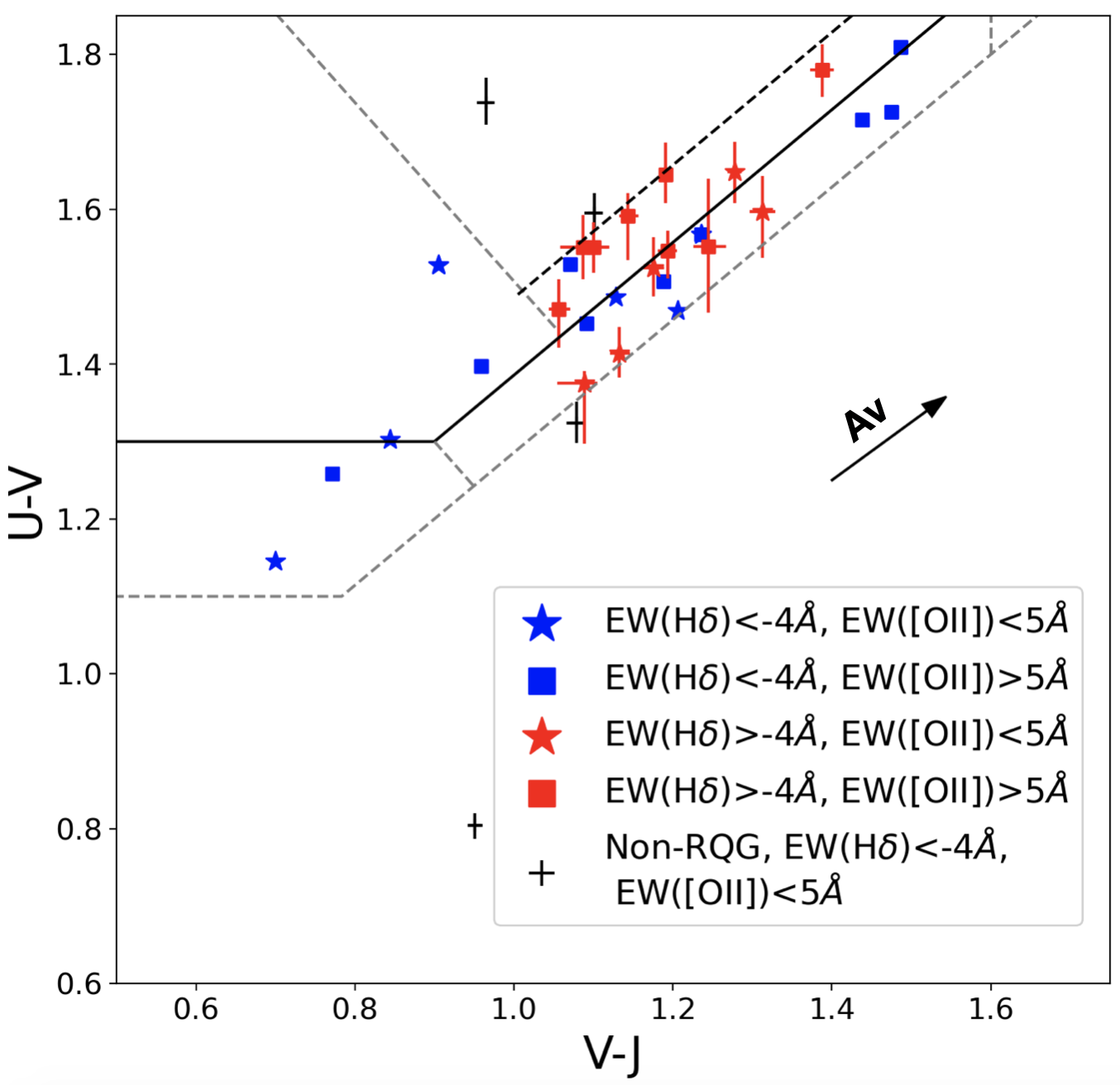}
    \caption{Blue marks are UVJ-selected RQGs with strong absorption lines. Red marks are UVJ-selected RQGs without strong absorption lines and their rest-frame UVJ colours' uncertainties. Star marks are galaxies without strong [OII] emission, while square marks are galaxies with strong [OII] emission lines. Cross marks are galaxies that are not selected as RQG on the UVJ diagram but with RQG spectral features and their UVJ colours' uncertainties. The arrow shows the reddening vector.}
    \label{fig:confirmation}
\end{figure}

The positions of these 27 galaxies on the UVJ diagram are plotted in Fig. \ref{fig:confirmation}. Among the 27 RQGs, 14 galaxies have strong Balmer absorption lines (EW(H$\delta$)<-4\AA). For the 13 galaxies below this threshold, we checked their colour index errorbars on the UVJ diagram (red marks in Fig. \ref{fig:confirmation}). Based on the fluxes and errors given in the COSMOS2015 catalogue, we randomly generate 1000 flux values following Gaussian distribution and then calculate rest-frame colour indices from these fluxes. In this way, we can calculate 3$\sigma$ errors for all galaxies. 9 galaxies out of 13 galaxies are close to the boundary. We also cross-matched these 13 galaxies with Chandra COSMOS Legacy Survey data \citep{Civano_2016}, and found two galaxies with H$\delta$ emission lines (ID 908883 and 635561) are X-ray AGNs. We should also notice that all these galaxies with weak Balmer absorption features lie in the slow quenching region, which confirms the quenching timescale of this UVJ region is longer, that their stellar populations have already grown old, and thus, the Balmer absorption feature started to fade. 6 of 13 galaxies (ID 314012, 375916, 391497, 438059, 559691, 810957, 816426, and 837551) have -4\AA\ <EW(H$\delta$)<-1.5\AA\ , meanwhile their spectra are passive. This is consistent with what we discuss in Appendix \ref{app:model track} Fig. \ref{fig:EWUVJ}. In Fig. \ref{fig:EWUVJ}, we show the evolution of EW along the evolutionary paths of model galaxies. The galaxies in slow quenching regions show weak H$\delta$ absorption lines (as weak as EW(H$\delta$)$\sim$1.5\AA). The remaining 3 galaxies are possible contamination. Therefore, we conclude that the overall contamination rate is not high.

\begin{figure}
	\includegraphics[width=\columnwidth]{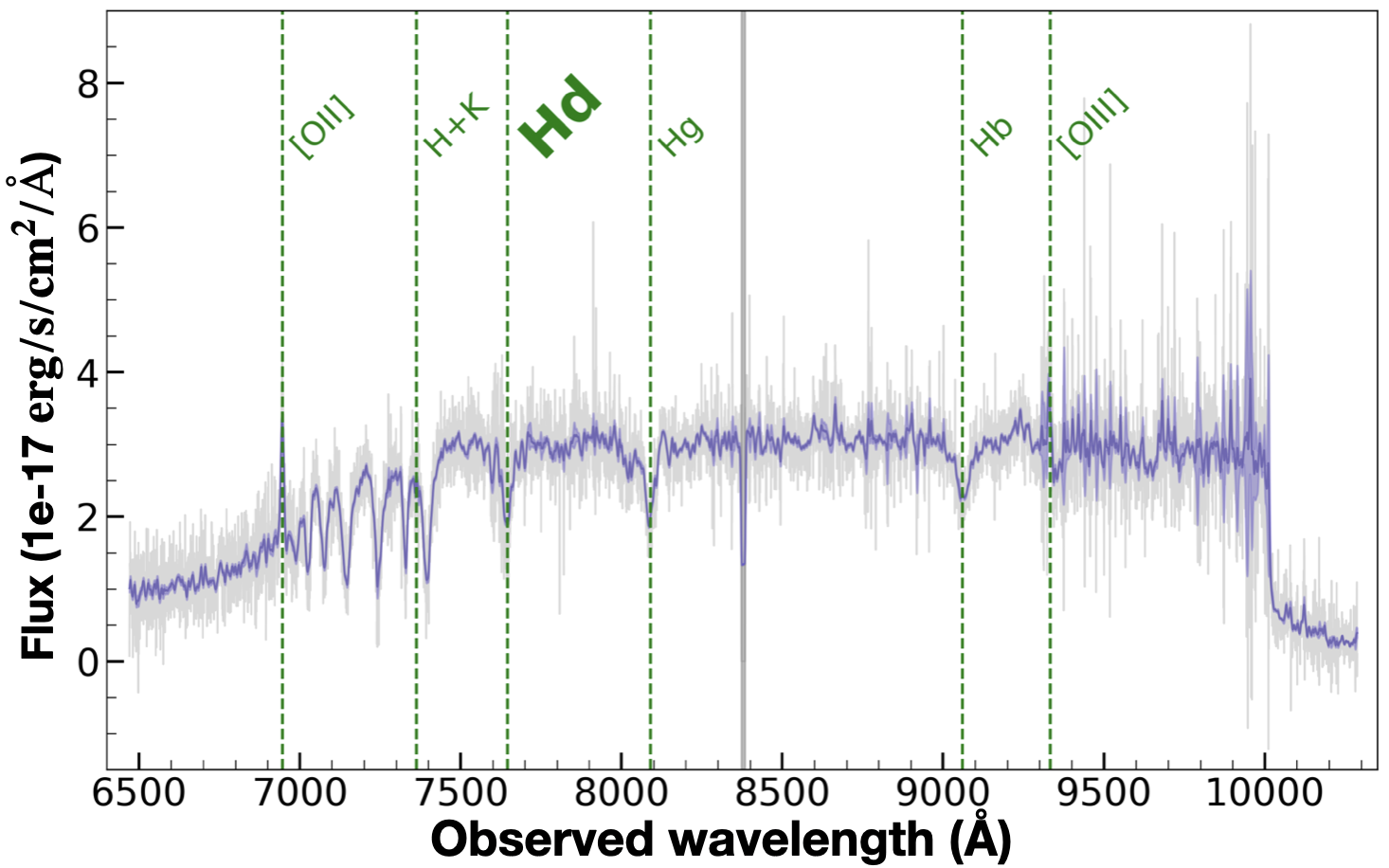}
	\includegraphics[width=\columnwidth]{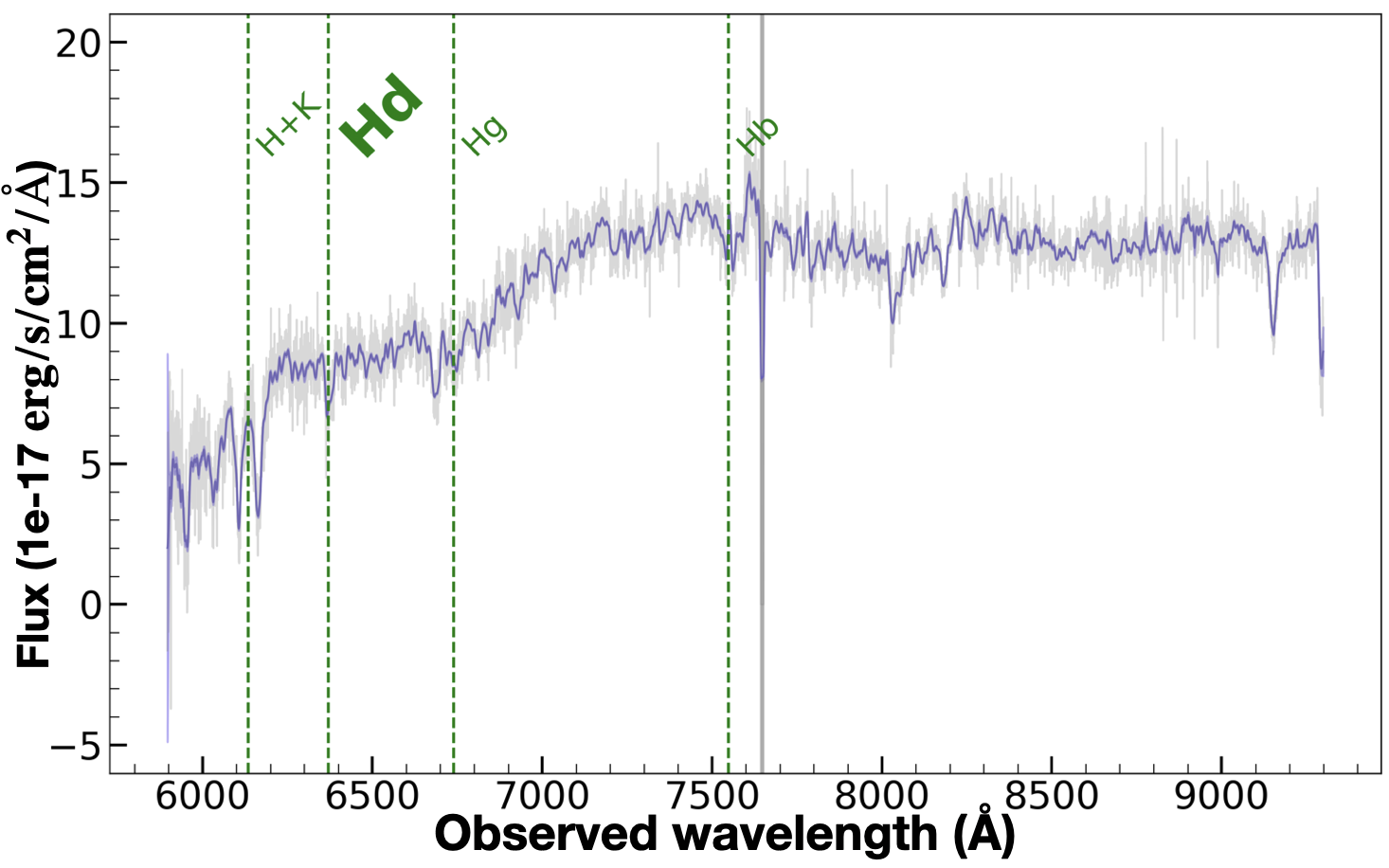}
    \caption{Two galaxy spectra (in observed frame) we checked. Spectral features are marked with green dashed lines. An artificial break between two chips of DEIMOS spectrograph is marked with grey shaded rectangular. {\bf Top panel}: A typical spectrum for strong Balmer absorption line galaxy, at z=0.864. {\bf Bottom panel}: A typical spectrum for weak Balmer absorption line galaxy, at z=0.553.}
    \label{fig:spectrum examples}
\end{figure}

However, in the 14 strong absorption line galaxies, 8 (6 of them are slow quenching galaxies) also have strong [OII] lines (EW([OII])>5\AA). This is consistent with the work of \citet{Rudnick_2017}, which finds [OII] emission in quenched galaxies. The [OII] emission may indicate that the depletion time is indeed longer for slow quenching galaxies, keeping the galaxy forming stars for a long time. However, we still need further work to fully understand the source of [OII] in RQGs. In Fig. \ref{fig:spectrum examples}, we present typical examples of one strong and one weak absorption galaxy spectra.

Besides the RQGs, we found 2 QGs and 2 SFGs that satisfy the RQG criteria in all high-quality spectra (Fig. \ref{fig:confirmation}). Their information is listed in Appendix \ref{app:spectra information}. The SFG on the bottom side (ID 897743) has clear AGN features (i.e. a broad MgII emission line, broad H$\gamma$ and H$\beta$ emission lines). The other SFG and one of the QGs are near the RQG region boundaries. However, for the other QG, which is far away from any boundary, we may have a risk of missing a certain subpopulation of RQG that, for some reason (e.g. undergoing stronger starburst before quenching), spend more time in the recently quenched phase.

In conclusion, our photometric UVJ selection method still suffers from certain contamination and completeness problems. But judging from the current spectroscopic data, neither the contamination nor incompleteness is high. Hence, our results from photometric data are still trustworthy. Due to the lack of high-quality spectra for absorption measurements, we do not have a statistical sample to estimate a reliable contamination or completeness rate. Deeper and wider spectroscopic observations of RQGs are required to better assess our photometric method. 

\subsection{Physical scenarios of quenching processes}\label{sec:mechanism}
In Sect. \ref{sec:qe}, we show that the quenching efficiency increases with both the stellar mass and the local density. From the quenching efficiency, we can see that the mass dependence trend is not greatly affected by the local density, but the efficiency is higher in denser local environments, especially in the highest local density bin. It shows that mass quenching happens in all kinds of environments, but the cluster environment (i.e. dense environment) can also enhance the quenching of massive galaxies. The most significant increase of efficiency appears in the most massive bin, indicating that mass-dependent quenching is much stronger for those massive galaxies. The quenching efficiency's dependence on local density is not independent. It is also affected by the stellar mass, so that for massive galaxies, this dependence almost disappears (see the middle and the right panels of Fig. \ref{fig:qesigma}). Therefore, although the environmental effects of quenching also exist, they are only efficient in affecting low mass systems. Our definition of quenching efficiency is different from commonly used definitions based on the fraction of quiescent galaxies (e.g. \citealt{Kawinwanichakij_2017, Vanderburg_2020, Reeves_2021}). Since our definition introduces the RQG population, which may still be in the process of quenching, it can convey the efficiency more directly. In \citet{Kawinwanichakij_2017}, the authors separate mass-dependent quenching efficiency and environmental quenching efficiency. To compare with their result, we also calculated the mass quenching efficiency using their definition (see equation 5 in \citealt{Kawinwanichakij_2017}). 
In \citet{Kawinwanichakij_2017}, the mass-dependent quenching efficiency is increasing with stellar mass, the trend is consistent with our result. In the least massive bin, our quenching efficiency is $\sim$0.3 dex higher, and in other bins, our efficiencies are $\sim$0.1 dex higher than \citet{Kawinwanichakij_2017} result. This might be due to different sample completeness or cosmic variance of our data.
The environmental quenching efficiency in \citet{Kawinwanichakij_2017} does not show a significant evolution with stellar mass. In our work, we only use two stellar mass bins while discussing environmental quenching, but instead, we discuss its dependence on local density. In general, low mass galaxies in our sample have lower efficiency, but this strongly depends on local density. We note that in \citet{Kawinwanichakij_2017}, the efficiency at the massive end has extremely large errorbars, therefore we can not establish a reliable comparison in that mass range.
In \citet{Vanderburg_2020}, the authors discuss quenching efficiency as a function of stellar mass, both the resulting trends and binned values agree with our result.
In \citet{Reeves_2021}, the authors discussed how the quiescent galaxy fraction depends on stellar mass. The trends they found agree with our results except in the least massive bin, where our quiescent galaxy value is slightly higher than theirs. This is likely a consequence of the different redshift range studied in their analysis ($1.0<z<1.5$). At that redshift, low mass galaxies are predominantly not quenched, hence their fraction is lower. The authors also pointed out that the environment and mass dependence of quenching are not separable, which is consistent with our result.
\citet{McNab_2021} studied the mass dependence of quenching by the excess of cluster SFG fraction over the field SFG fraction. The increasing trend of the excess represented quenching efficiency agrees with our result. \citet{Webb_2020} studied the SFR declination rate in field and cluster galaxies by spectroscopy and photometry fitting. Their results show cluster galaxies (i.e. denser environment galaxies) quench faster, which also agrees with our results.

In Sect. \ref{sec:qs}, we see the quenching stage decreases with both the stellar mass and the local density. The quenching stage provides us with a glimpse of quenching in the past. It significantly decreases with stellar mass in the low mass end (similarly for the cluster candidates and the field), showing that more massive galaxies are in a much later stage of quenching. This is consistent with the downsizing scenario (e.g. \citealt{Cowie_1996,Cattaneo_2008,Webb_2020}) that massive galaxies form and quench earlier, hence they are in a later, more advanced stage of evolution compared to low mass galaxies. The mass dependence of the quenching stage is strongly affected by the local density. The trend is only significant in high-density environments. For sparse environments, the errorbars are too large for us to say anything. The scenario is consistent with \citet{Hatch_2011}, where faster evolution in protocluster is found. This can also be due to the migration of galaxies from low density cluster outskirts to high density cluster cores where the quenching happens. The quenching stage also shows a stable decreasing trend with the local density, indicating that galaxies in denser environments (cluster cores) are in a later, more advanced stage of quenching. The earlier quenching stage of field galaxies also supports this scenario. This is consistent with the inside-out quenching in clusters, where galaxies in cluster cores quench early, while the galaxies in the outer parts quench progressively later, this is consistent with the scenario in \citet{Koyama_2013, Shimakawa_2018}. Another possibility is that during the in-falling process, their gas will be stripped off, which directly leads to their quenching \citep{Bekki_2002}. In this case, the effect should be stronger for low mass galaxies, since their gas components are easier to strip. This environmental dependence is not strongly affected by the stellar mass, which supports the inside-out quenching that has an impact on low and high mass galaxies equally. In \citet{McNab_2021}, the authors use three types of transitional populations, green valley (GV) selected by rest-frame (NUV-V) and (V-J) colours, blue quiescent (BQ) selected by rest-frame (U-V) and (V-J) colours, and post-starburst (PSB) selected by galaxy spectra. They connect the quenching timescale to the transitional population fraction excess in clusters. They found that the lowest mass systems ($<10^{10.5}M_{\odot}$) have short quenching timescales (<1 Gyr), which agrees with our result that low mass galaxies are dominated by fast quenching (<1 Gyr).

The slow over fast quenching ratio is an indicator of the quenching timescale. In Sect. \ref{sec:timescale}, we find the massive galaxies are dominated by slow quenching, and the less massive galaxies are dominated by fast quenching. This result is consistent with \citet{Walters_2022}, where the authors found slow quenchers tend to have higher stellar mass. The authors used IllustrisTNG simulation \citep{Nelson_2019} to estimate the quenching timescale. \citet{Walters_2022} also used MaNGA IFU survey data \citep{Bundy_2015} and took measurements of stellar age gradient from \citet{Woo_2019} to split fast and slow quenchers. In this way, the authors confirmed that slow quenchers have higher stellar mass. 
In \citet{Hahn_2017}, the authors matched N-body simulation from \citet{Wetzel_2013} and their evolutionary model with SDSS DR7 \citet{Abazajian_2009} data to investigate and constrain the quenching timescale. They find that satellite galaxies have shorter quenching timescales than the central ones. This result broadly agrees with this work, assuming that satellite galaxies are less massive compared to central galaxies. However, we note that the aperture we use for cluster member galaxy selection in this work is about a factor of 2 larger than $R_{200}$ in our clusters (see \ref{tab:X-ray cluster} and $R_{200}$ in Table 2 of \citealt{Gozaliasl_2019}). Therefore we cannot exclude the possibility that some galaxies far from the cluster centre could also be considered central galaxies of smaller haloes infalling towards the main cluster core.

In Sect. \ref{sec:cases}, two specified cases are discussed. These two cases show clearly different timescales of quenching. The mass-dependent quenching in sparse regions should be less affected by the environment. It turned out to have a longer timescale, and very likely to be AGN feedback that caused quenching. This is consistent with the observed timescale of quenching for this process \citep{Hopkins_2006} and with the typical stellar mass of host galaxies \citep{Kauffmann_2003}. AGN feedback also plays an indirect role in the morphology transformation by suppressing in situ star formation, and AGN feedback is essential for forming ellipticals \citep{Dubois_2016}. This could also explain the bimodality in galaxy morphology. The environmental quenching is only significant for low mass galaxies, and it is surely stronger in denser regions. Quenching, in this case, appears to be a short timescale type, which we reasonably speculate to be caused by ram pressure stripping. As mentioned in Sect. \ref{sec:cluster catalogue}, three cluster candidates are detected in X-ray, and their total masses are measured, which indicates that they are massive enough to allow ram pressure stripping to happen. 

\begin{figure}
	\includegraphics[width=\columnwidth]{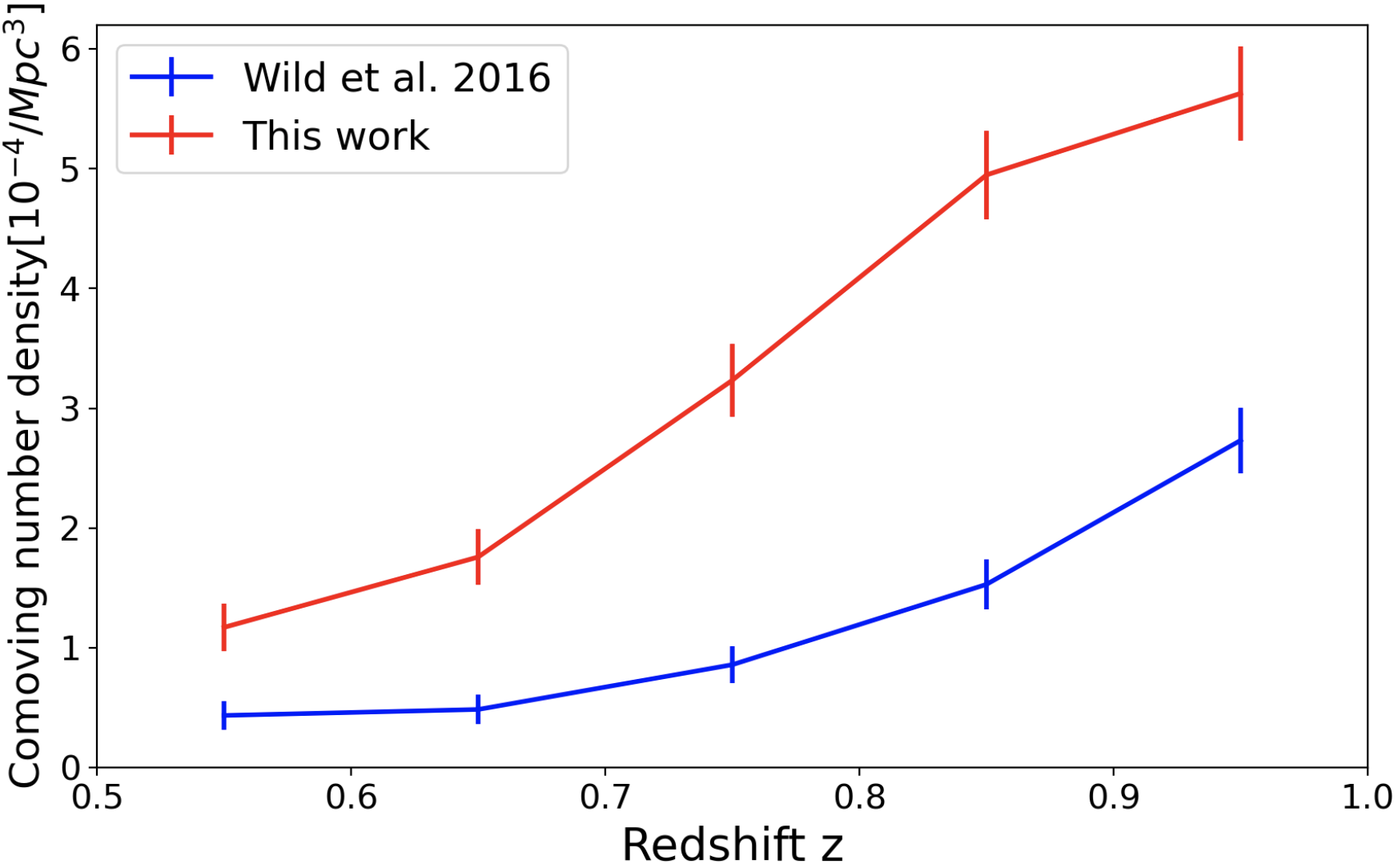}
    \caption{A comparison between comoving number density of this work (RQG in traditional UVJ selection region) and \citet{Wilkinson_2021} (PSB) in COSMOS field. The five redshift bins in this plot are 0.5-0.6, 0.6-0.7, 0.7-0.8, 0.8-0.9, and 0.9-1.0.}
    \label{fig:density compare}
\end{figure}

In Sects. \ref{sec:timescale} and \ref{sec:number density}, we had a preliminary explore the quenching's dependence on redshift. In both cluster candidates and the general field, the fast quenching RQGs are increasing with redshift. In cluster candidates, slow quenching RQG over fast quenching RQG ratio is decreasing with redshift, indicating that fast quenching becomes more important at higher redshift. In \citet{Tinker_2010,Balogh_2016, Fossati_2017, Foltz_2018}, the authors discussed galaxy quenching timescale's dependence on redshift in clusters and groups. Combining their results, the quenching timescale also decreases with redshift.
We estimated the comoving number density of fast quenching and slow quenching RQGs in the field. Fast quenching RQGs show a strong increasing trend. This trend is consistent with post-starburst analysis in \citet{Wild_2016} and \citet{Belli_2019}. The redshift range they investigate is wider than ours, but both show that post-starburst galaxies' number density is increasing with redshift up to z$\sim$2. 
\citet{Wild_2016} analyses the PSBs in UKIDSS UDS field. To compare with our results, we use the PSB catalogue from \citet{Wilkinson_2021} that includes PSBs from the COSMOS field. To make a direct comparison, we limit our RQG sample to the traditional PSB region in \citet{Wild_2014} (region 1 in Fig. \ref{fig:UVJ}). We also apply the same mass cut for both samples. Then we compare the comoving number density of two samples in Fig. \ref{fig:density compare}. The increasing trend is found in both samples, while the absolute value has $\sim$0.4 dex difference, especially in higher redshift bins. We note that \citet{Wilkinson_2021} Fig. 1 shows a spectroscopic confirmation of their PSB sample, where a large fraction of PSBs are missed by their PCA selection. The incompleteness of PCA selection may lead to the difference between the two samples.
 
In this work, slow quenching RQGs also show an increasing trend, although it is much weaker. This indicates that fast quenching processes become more important at higher redshifts. At lower redshifts, pre-processing, which is also a long timescale mechanism \citep{Taranu_2014}, may be an additional quenching mechanism. This may not be as common at high redshifts because not many dense groups would exist. Therefore, the growing influence of pre-processing can result in the rising importance of slow quenching at lower redshift.

\bigskip
In conclusion, our photometric selection method is also a powerful tool in determining target galaxies for large systematic spectroscopic surveys of RQGs. As we apply this method to COSMOS galaxies in cluster candidates at 0.5<z<1.0, we obtain results with high consistency with previous works. However, our statistical analysis still suffers from the limited sample size and the uncertainty introduced by the photometric selection. We plan to increase our sample by including all Hyper Suprime-Cam Subaru Strategic Program (HSC-SSP) DEEP fields \citep{Aihara_2019}. If we include IRAC data, we can also expand our sample redshift range from the current 0.5-1.0 to 0.5-2.0, which traces the quenching process back to the cosmic noon. More importantly, spectroscopic confirmation of our sample's RQG features is required for further analysis. The preliminary spectroscopic confirmation shows low contamination and high completeness, but more spectra are required for more accurate confirmation. The high accuracy contamination and completeness rates of our UVJ selection method must be estimated and corrected to get the final results. Moreover, RQG galaxy spectra can be used to constrain quenching timescales through the spectral fitting. In this way, we can better distinguish between different mechanisms. We look forward to the Prime Focus Spectrograph (PFS) \citep{Tamura_2018} on Subaru that will come into use in 2023, when we can efficiently obtain the spectra of a large sample of RQGs due to the PFS's large FoV (1.3 deg$^2$) and great multiplicity (2,400 fibres).

\section{Conclusions}
In this study, we construct a multiband photometric sample of galaxies from 17 cluster candidates in the COSMOS field at $0.5<z<1.0$. We then adopt the rest-frame UVJ selection method to classify recently-quenched, star-forming and quiescent galaxies. Moreover, we apply two exponentially declining SFH models with different quenching timescales to separate fast quenching and slow quenching galaxies on the UVJ diagram. We then correct for the projection effects introduced by the uncertainties of photometric redshifts and investigate different quenching scenarios as a function of stellar mass and local density. The main results of this study are summarised as follows:
\begin{enumerate}
\item We define quenching efficiency as the ratio between the number of RQG and SFG in a given volume. Quenching efficiency increases with both mass and local density, confirming the existence of mass and environment-dependent quenching. Mass-dependent quenching is stronger for high mass galaxies.
\item We define the quenching stage as the ratio between RQG and QG, aiming to quantify the history of quenching in a given volume. We find that this parameter also depends on both mass and local density. Massive galaxies are in a later stage of the quenching process, which is consistent with the downsizing scenario. Galaxies in the cluster core regions are in an earlier stage of quenching, which is a sign of inside-out quenching in cluster environments.
\item We compare the quenching efficiency and quenching stage of cluster and field galaxies. We find that mass quenching is enhanced by the cluster environment, and the field galaxies are in an earlier quenching stage compared to the cluster galaxies of the same stellar mass.
\item The ratio between the number of slow and fast quenching galaxies is a proxy of the quenching timescale. In cluster environments, the quenching timescale depends both on the stellar mass and redshift.
Low-mass galaxies are dominated by fast quenching, while massive galaxies are dominated by slow quenching. Slow quenching is also more important at lower redshifts.
\item Massive galaxies' quenching in the sparse environment is dominated by the slow quenching process. This is likely due to the relatively long timescale of AGN feedback, as AGNs are hosted exclusively by massive galaxies.
\item Fast quenching is much more important to quench low-mass galaxies in the dense environment. We believe this is likely caused by short timescale processes such as ram pressure stripping, which only happens close to the cluster core (i.e. high-density environment) and affects low mass galaxies more efficiently.
\item  Visibility time of RQGs may degenerate with the quenching properties, especially with quenching efficiency and timescale. However, we cannot separate well the visibility time and the quenching properties without spectroscopic data. We leave this problem for our future spectroscopic work.
\item Existing spectroscopic data confirms that our photometry-based UVJ selection method is reliable enough. By measuring the EW of Balmer absorption lines and [OII] emission line from DEIMOS 10K spectroscopic survey data, we can confirm the RQG nature of a limited number of galaxies with high-quality spectra. Although 7 UVJ-selected RQGs do not show expected strong absorption lines, 2 of them are AGN host galaxies, 4 of them lie within the slow quenching RQG region and show relatively weak absorption lines. Thus, only one of the 17 galaxies cannot be explained, and it is likely contamination. Moreover, only 3 galaxies outside the RQG regions have RQG spectral features, and two of them are very close to the selection boundary. Combining all these pieces of evidence, the UVJ selection method is a powerful and reliable analysis tool to investigate the RQG population before more spectroscopic data become available.
\end{enumerate}

The results we present in this work give evidence of both mass-dependent quenching and environmental quenching. They also shed light on the mass and environmental dependence of quenching timescales. With future spectroscopic data, the physical properties of our sample galaxies can be better constrained, and the uncertainty of current photometric results can be reduced. The UVJ selection method we displayed in this paper can be applied to large photometric datasets (e.g. HSC-SSP data) of optical to NIR. We can efficiently select RQG candidates for further spectroscopic surveys performed by high multiplicity multiobject instruments such as PFS. 

\begin{acknowledgements}
      We thank the anonymous referee for his/her helpful comments, which contributed to improving the paper. We thank Drs.\ Omar Almaini and Sirio Belli for helpful discussions. We also thank Dr. Omar Almaini and his colleagues for kindly sharing PSB catalogue in COSMOS field selected by PCA method. TK acknowledge the financial support by JSPS Kakenhi (\#18H03717).Two of us (ZM, NY) acknowledge support from GP-PU at Tohoku University and JST, the establishment of university fellowships towards the creation of science technology innovation, Grant Number JPMJFS2102.
\end{acknowledgements}

\bibliographystyle{aa}
\bibliography{example}
%
%

\begin{appendix}
\section{Quenching model tracks on the rest-frame UVJ diagram}\label{app:model track}
{
In this part, we investigate the evolution tracks of model galaxies in the rest-frame UVJ space and discuss how they depend on each parameter. 

\smallskip
First, we discuss how the redshift might influence the UVJ selection criteria. In Fig. \ref{fig:redshiftUVJ}, we show model galaxies with different star formation duration. The time after quenching started is marked with stars in different colours. We can see the evolution path is almost the same in the RQG and QG regions, indicating that a star formation duration difference of 3 Gyr will not change the selection criteria of RQGs. Suppose two galaxies started to form at the same time, but have two different duration of star formation of 1 and 4 Gyr. They will show almost the same colours if the times after the start of quenching are the same. The observed epochs of these two RQGs can, for example, correspond to $z$=1.0 and 0.5, respectively. Therefore, we do not need to shift the selection boundaries depending on the redshifts in this range. 
\begin{figure}
	\includegraphics[width=\columnwidth]{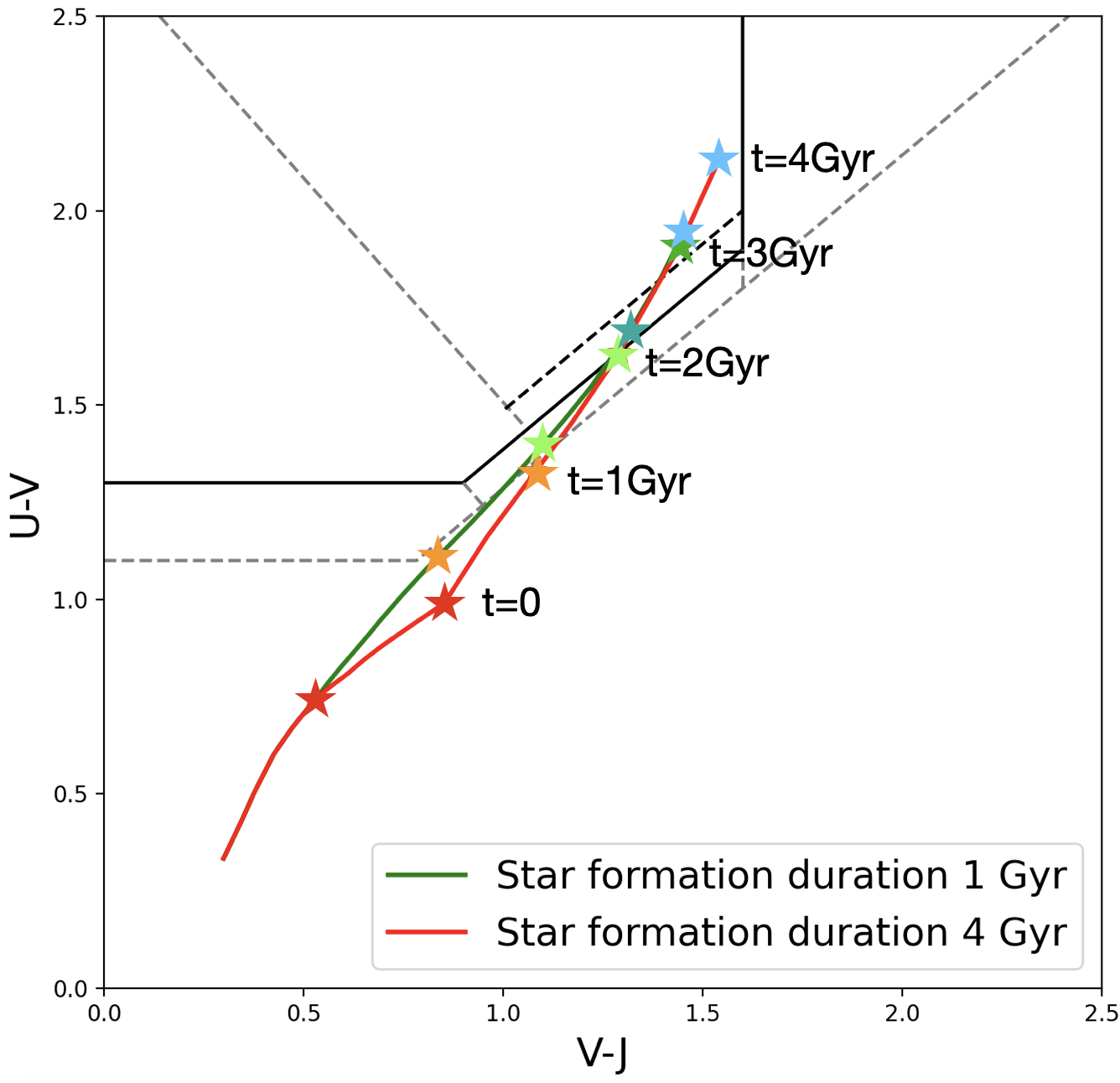}
    \caption{This plot includes galaxies with different star formation duration times in SFHs (see Fig. \ref{fig:SFH}). The green line with a duration of 1 Gyr, and the red line with a duration of 4 Gyr. Both with a quenching timescale of 1 Gyr. Star marks show the time after the start of quenching. The star marks with the same colour on two curves show the same time after quenching.}
    \label{fig:redshiftUVJ}
\end{figure}

\smallskip
Then we discuss the evolution of spectral features along the model paths. We use the high-resolution mode of CIGALE to output high-resolution SED of model galaxies at each position and calculate the equivalent widths (EWs) of H$\delta$ and [OII] line features. Figure \ref{fig:EWUVJ} shows the fast and slow quenching paths with blue and red curves, respectively. The EWs of H$\delta$ (black) and [OII] (light blue) at each point are presented next to the position. Negative EW represents the absorption feature, while positive EW represents the emission feature. All EWs are calculated using the same method as in Sect. \ref{sec:confirmation}. We can see that the H$\delta$ absorption line strength first increases and then decreases along the evolution paths. In the RQG region, both the fast quenching and slow quenching galaxies show strong H$\delta$ absorption. For the fast quenching galaxy, there is no [OII] emission left. However, the H$\delta$ absorption is weaker for the slow quenching galaxy compared to the fast quenching galaxy due to its longer quenching timescale, and the [OII] emission is stronger due to remaining star formation. These features are also consistent with the spectral confirmation in Sect. \ref{sec:confirmation}. In conclusion, our boundary selection is reasonable and consistent with the spectral features of RQGs.

\begin{figure}
	\includegraphics[width=\columnwidth]{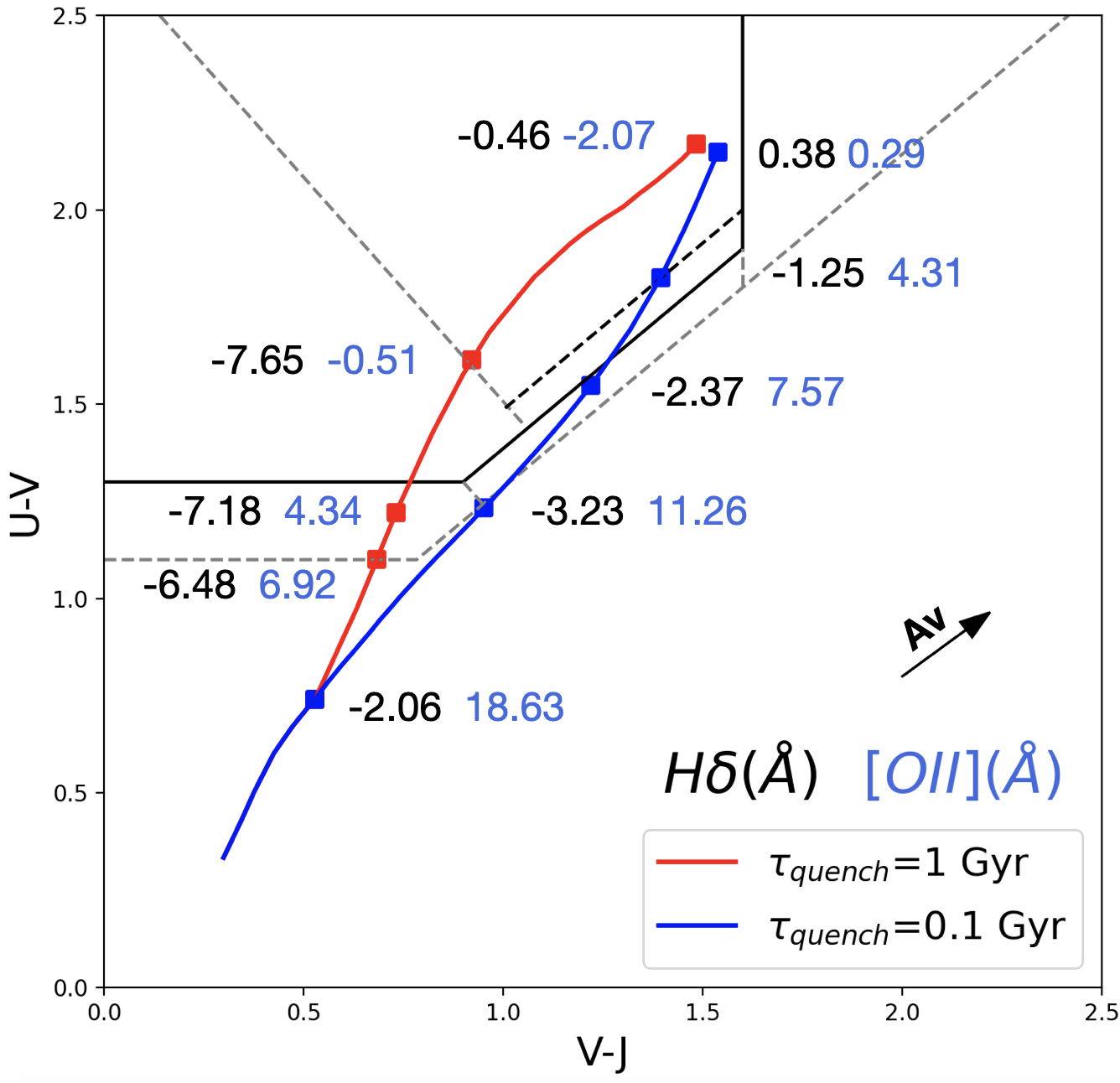}
    \caption{This plot shows the equivalent width (EW) of H$\delta$ absorption line and [OII] emission line at different positions in UVJ space. Two evolutionary tracks are of galaxies with 1 Gyr star formation duration but different quenching timescale, E(B-V)=0.15. Values in blue show EW of [OII] emission lines, values in black EW of H$\delta$ absorption lines. A negative value represents absorption, and a positive value represents emission.}
    \label{fig:EWUVJ}
\end{figure}

\smallskip
Finally, we discuss how dust attenuation affects the evolution paths. We use different dust parameters in CIGALE and calculate the evolution paths on the UVJ diagram. In Fig. \ref{fig:dustUVJ}, we show evolution path with E(B$-$V)=0.15 and 0.2. The paths are shifted to the top right, and the reddening vector of E(B$-$V)=0.1 is also plotted on the diagram. We note that for both E(B$-$V) values, the fast quenching and slow quenching tracks are still in the fast (purple) and slow (orange) quenching regions. Considering RQGs should have a certain amount of dust as the fuel of the remaining star formation, the dust attenuation will lead to some contamination but will not affect the separation of fast and slow quenching much.
\begin{figure}
	\includegraphics[width=\columnwidth]{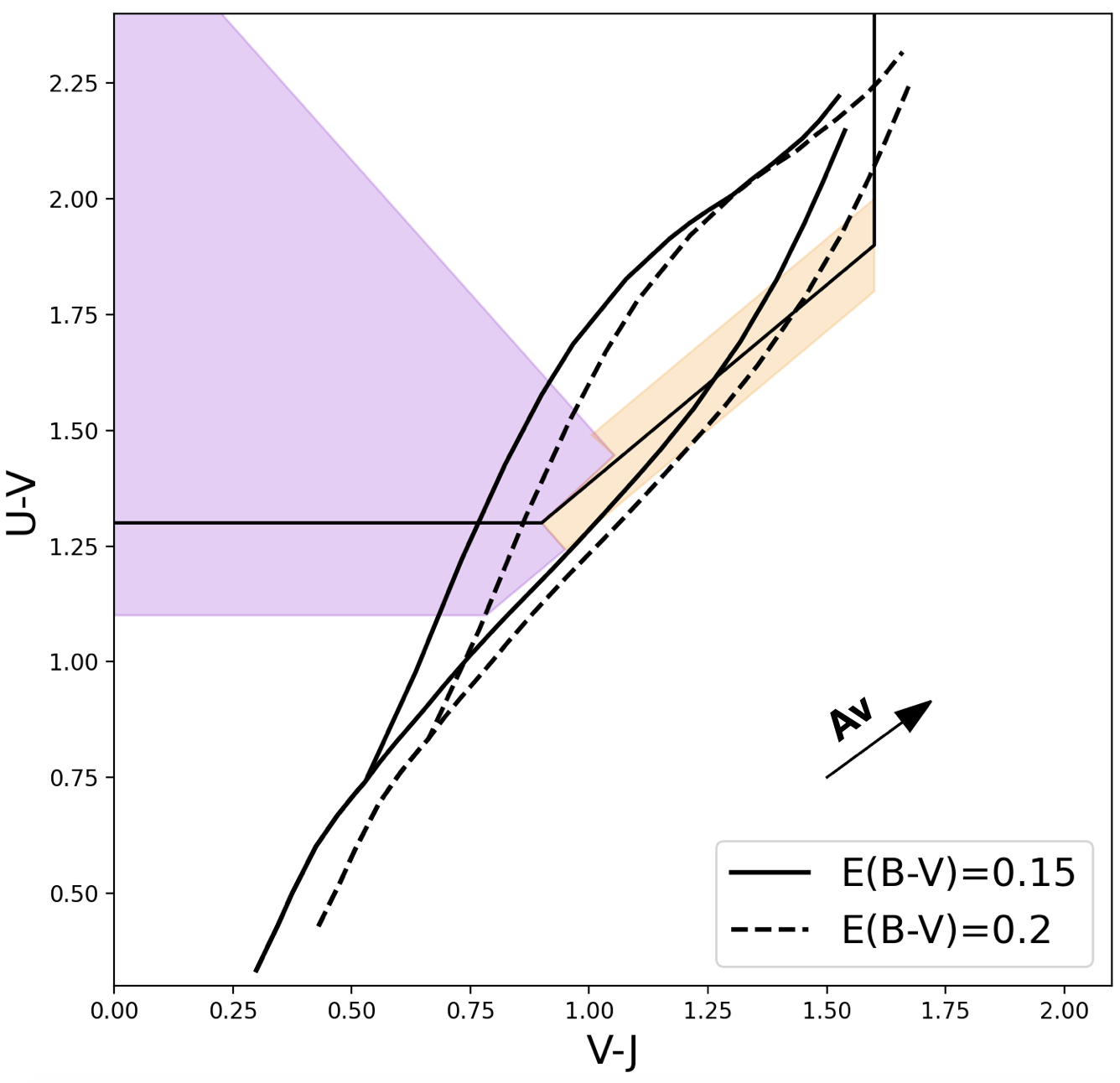}
    \caption{This plot shows the effect of dust attenuation. Solid tracks are paths with E(B-V)=0.15 dust attenuation. Dashed tracks are  E(B-V)=0.2 dust attenuation. The arrow on the plot shows the reddening vector of E(B-V)=0.1.}
    \label{fig:dustUVJ}
\end{figure}
}

\section{Binning information for result}\label{app:bin information}
In this part, we introduce the bin information of all bins we used in Sect. \ref{sec:result}. The first columns of all tables show the ranges of bins, and the second columns show the numbers of galaxies in these bins.
\begin{table}
\centering
\caption{Information of mass bins in all environments (left panel in Fig. \ref{fig:qemass}, \ref{fig:qsmass}, and \ref{fig:fast slow separated}).}
\setlength{\tabcolsep}{1.7mm}{
\begin{tabular}{ccc}
\hline
\\[-3mm]
Mass range [$log[M/M_{\odot}$] & number of  & number of\\
& cluster galaxies & field galaxies
\\[1mm]
\hline
\\[-3.3mm]
9.80-10.10&349&5027\\
10.10-10.42&346&4127\\
10.42-10.74&351&3450\\
10.74-&350&3296\\
\\[-3.3mm]
\hline\\
\end{tabular}}
\label{tab:massbin}
\end{table}

\begin{table}
\centering
\caption{Information of local density bins with all mass (left panels in Fig. \ref{fig:qesigma} and \ref{fig:qssigma}, and right panels of Fig. \ref{fig:fast slow separated}).}
\setlength{\tabcolsep}{1.7mm}{
\begin{tabular}{cc}
\hline
\\[-3mm]
Local density range [$/Mpc^2$] & number of galaxies\\
\\[-3mm]
\hline
\\[-3.3mm]
field&15900\\
0.40-1.02&347\\
1.02-1.30&345\\
1.30-1.67&355\\
1.67-&349\\
\\[-3.3mm]
\hline\\
\end{tabular}}
\label{tab:sigmabin}
\end{table}

\begin{table}
\centering
\caption{Information of mass bins in low local density (middle panel in Figs. \ref{fig:qemass} and \ref{fig:qsmass}) (top table), and of mass bins in high local density (right panel in Fig. \ref{fig:qemass} and \ref{fig:qsmass}) (bottom table).}
\setlength{\tabcolsep}{1.7mm}{
\begin{tabular}{cc}
\hline
\\[-3mm]
Mass range [$log[M/M_{\odot}$] & number of galaxies\\
\\[-3mm]
\hline
\\[-3.3mm]
9.80-10.16&231\\
10.16-10.54&231\\
10.54-&230\\
\\[-3.3mm]
\hline\\
\end{tabular}}

\setlength{\tabcolsep}{1.7mm}{
\begin{tabular}{cc}
\hline
\\[-3mm]
Mass range [$log[M/M_{\odot}$] & number of galaxies\\
\\[-3mm]
\hline
\\[-3.3mm]
9.80-10.24&236\\
10.24-10.69&237\\
10.69-&231\\
\\[-3.3mm]
\hline\\
\end{tabular}}
\label{tab:massbin2}
\end{table}

\begin{table}
\centering
\caption{Information of mass bins in low local density (middle panel in Fig. \ref{fig:qesigma} and \ref{fig:qssigma}) (top table), and of mass bins in high local density (right panel in Fig. \ref{fig:qesigma} and \ref{fig:qssigma}) (bottom table).}
\setlength{\tabcolsep}{1.7mm}{
\begin{tabular}{cc}
\hline
\\[-3mm]
Local density range [$/Mpc^2$] & number of galaxies\\
\\[-3mm]
\hline
\\[-3.3mm]
0.40-1.02&233\\
1.02-1.49&231\\
1.49-&231\\
\\[-3.3mm]
\hline\\
\end{tabular}}

\setlength{\tabcolsep}{1.7mm}{
\begin{tabular}{cc}
\hline
\\[-3mm]
Local density range [$/Mpc^2$] & number of galaxies\\
\\[-3mm]
\hline
\\[-3.3mm]
0.40-1.16&233\\
1.16-1.57&235\\
1.57-&233\\
\\[-3.3mm]
\hline\\
\end{tabular}}
\label{tab:sigmabin2}
\end{table}

\begin{table}
\centering
\caption{Top table: mass bin information in Fig. \ref{fig:fsmass}. Middle table: local density bin information in Fig. \ref{fig:fssigma}. Bottom table: redshift bin information in Fig. \ref{fig:fsz}.}
\setlength{\tabcolsep}{1.7mm}{
\begin{tabular}{cc}
\hline
\\[-3mm]
Mass range [$log[M/M_{\odot}$] & number of galaxies\\
\\[-3mm]
\hline
\\[-3.3mm]
9.80-10.25&82\\
10.25-10.71&82\\
10.71-&82\\
\\[-3.3mm]
\hline\\
\end{tabular}}

\setlength{\tabcolsep}{1.7mm}{
\begin{tabular}{cc}
\hline
\\[-3mm]
Local density range [$log /Mpc^2$] & number of galaxies\\
\\[-3mm]
\hline
\\[-3.3mm]
0.52-1.14&83\\
1.14-1.50&81\\
1.50-&82\\
\\[-3.3mm]
\hline\\
\end{tabular}}

\setlength{\tabcolsep}{1.7mm}{
\begin{tabular}{cc}
\hline
\\[-3mm]
Redshift range & number of galaxies\\
\\[-3mm]
\hline
\\[-3.3mm]
0.50-0.70&80\\
0.70-0.86&83\\
0.86-1.00&83\\
\\[-3.3mm]
\hline\\
\end{tabular}}
\label{tab:binz}
\end{table}

\section{Detailed information of spectral confirmation}\label{app:spectra information}
In this part, we list the information of galaxy spectra in Sect. \ref{sec:confirmation} Fig. \ref{fig:confirmation} in Table \ref{tab:spec_info}. All IDs are the IDs of galaxies in \citet{Laigle_2016}.

\begin{table*}
\caption{Information of galaxy spectra in Sect. \ref{sec:confirmation} Fig. \ref{fig:confirmation}. The first column gives the source of the spectroscopic data. The Second column shows galaxy ID. The third and fourth columns show rest-frame UVJ colours. The fifth and sixth columns show equivalent width of H$\delta$ and [OII] lines; a positive value means the emission and a negative value means an absorption. The seventh column shows the type of galaxy classified by rest-frame UVJ colours.}
\centering
\begin{tabular}{cccccccc}
\hline
\\[-3mm]
Dataset & ID & V-J & U-V & H$\delta$[\AA] & [OII][\AA] & Type & Comment\\
\\[-3mm]
\hline
\\[-3.3mm]
DEIMOS &246749 & 1.13 & 1.49 & -5.74$\pm$0.19 & 1.80$\pm$0.27 & RQG &\\
DEIMOS &845010 & 0.70 & 1.15 & -5.67$\pm$0.75 & / & RQG &\\
DEIMOS &228878 & 1.21 & 1.47 & -5.12$\pm$0.49 & 4.76$\pm$0.75 & RQG &\\
DEIMOS &240895 & 0.84 & 1.30 & -4.31$\pm$0.07 & 2.81$\pm$0.17 & RQG &\\
DEIMOS &758494 & 0.77 & 1.26 & -9.42$\pm$0.23 & 33.77$\pm$0.20 & RQG &\\
DEIMOS &402222 & 0.96 & 1.40 & -6.69 $\pm$0.01 & 9.41$\pm$0.02 & RQG &\\
DEIMOS &255316 & 1.07 & 1.53 & -6.64$\pm$0.07 & 7.59$\pm$0.12 & RQG &\\
DEIMOS &610145 & 1.09 & 1.45 & -5.42$\pm$0.22 & 10.31$\pm$0.35 & RQG &\\
DEIMOS &631292 & 1.19 & 1.51 & -4.35$\pm$0.28 & 6.10$\pm$0.46 & RQG &\\
DEIMOS &609077 & 1.49 & 1.81 & -4.54$\pm$0.12 & 14.82$\pm$0.28 & RQG &\\
DEIMOS &559691 & 1.19 & 1.65 & -3.87$\pm$0.10 & 0.77$\pm$0.19 & RQG &\\
DEIMOS &375916 & 1.31 & 1.60 & -2.06$\pm$0.02 & / & RQG &\\
DEIMOS &391497 & 1.09 & 1.38 & -1.81$\pm$0.08 & 4.75$\pm$0.06 & RQG &\\
DEIMOS &908883 & 1.13 & 1.49 & 16.25$\pm$0.79 & / & RQG & X-ray AGN\\
DEIMOS &314012 & 1.13 & 1.41 & -3.50$\pm$0.13 & 34.46$\pm$0.69 & RQG &\\
DEIMOS &745276 & 1.24 & 1.55 & -0.58$\pm$0.35 & 17.67$\pm$0.75 & RQG &\\
DEIMOS &635561 & 1.10 & 1.55 & 4.08$\pm$0.06 & 8.44$\pm$0.13 & RQG & X-ray AGN\\
DEIMOS &897743 & 0.95 & 0.80 & -4.32$\pm$0.34 & 2.89$\pm$0.32 & SFG & AGN\\
DEIMOS &855876 & 1.08 & 1.32 & -4.51$\pm$0.06 & 4.01$\pm$0.11 & SFG &\\
DEIMOS &353441 & 0.97 & 1.74 & -5.53$\pm$0.15 & 3.26$\pm$0.60 & QG &\\
DEIMOS &816527 & 1.10 & 1.60 & -6.39$\pm$0.03 & 3.24$\pm$0.06 & QG &\\
LEGA-C&835004&0.90&1.53&-6.68$\pm$0.34&4.36$\pm$0.09&RQG&\\
LEGA-C&777794&1.44&1.72&-6.42$\pm$0.44&12.64$\pm$0.11&RQG&\\
LEGA-C&811857&1.24&1.57&-4.24$\pm$0.68&/ &RQG&\\
LEGA-C&837551&1.39&1.78&-3.53$\pm$0.62&-0.40$\pm$0.03&RQG&\\
LEGA-C&816426&1.06&1.47&-3.17$\pm$0.35&/&RQG&\\
LEGA-C&438059&1.14&1.59&-3.03$\pm$0.61&4.37$\pm$0.03&RQG&\\
LEGA-C&810957&1.19&1.55&-1.52$\pm$0.37&38.56$\pm$0.36&RQG&\\
LEGA-C&865973&1.09&1.55&-1.26$\pm$0.79&13.33$\pm$0.51&RQG&\\
LEGA-C&621801&1.18&1.52&-1.20$\pm$0.48&0.08$\pm$0.07&RQG&\\
LEGA-C&810435&1.47&1.73&-4.15$\pm$0.65&28.88$\pm$0.09&RQG&\\
\\[-3.3mm]
\hline\\
\end{tabular}
\label{tab:spec_info}
\end{table*}

\end{appendix}

\end{document}